\ifdefined\pdfminorversion\pdfminorversion=7\fi
\documentclass{article} %
\usepackage{iclr2027_conference,times}

\usepackage{amsmath,amsfonts,bm}

\newcommand{\captiona}{{\em (a)}}
\newcommand{\captionb}{{\em (b)}}
\newcommand{\captionc}{{\em (c)}}
\newcommand{\captiond}{{\em (d)}}

\def\eqref#1{equation~\ref{#1}}

\def\eps{{\epsilon}}

\def\vf{{\bm{f}}}
\def\vk{{\bm{k}}}
\def\vu{{\bm{u}}}
\def\vx{{\bm{x}}}

\newcommand{\uvel}{\vu}                         %
\newcommand{\pos}{\vx}                          %
\newcommand{\uhat}{\hat{\vu}}                   %
\newcommand{\vort}{\bm{\omega}}                 %
\newcommand{\press}{p}                          %
\newcommand{\visc}{\nu}                         %
\newcommand{\force}{\vf}                        %
\newcommand{\domside}{L_0}                      %

\newcommand{\urms}{u'}                          %
\newcommand{\enst}{\Omega}                      %
\newcommand{\diss}{\eps}                        %
\newcommand{\intscale}{L}                       %
\newcommand{\Reyn}{\mathit{Re}}                 %
\newcommand{\Relam}{\mathit{Re}_\lambda}        %

\newcommand{\wave}{\vk}                         %
\newcommand{\kmaxi}{k_{\max}}                   %

\newcommand{\dx}{\Delta x}                      %
\newcommand{\dtsolv}{\Delta t}                  %
\newcommand{\dtsnap}{\Delta t_{\mathrm{snap}}}  %

\usepackage{hyperref}
\usepackage{url}
\newcommand{\doi}[1]{doi: \href{https://doi.org/#1}{\begingroup
  \urlstyle{rm}\renewcommand{\UrlFont}{\fontencoding{T1}\rmfamily\selectfont}%
  \nolinkurl{#1}\endgroup}}
\usepackage{graphicx}
\usepackage{wrapfig}
\graphicspath{{./}}
\usepackage{booktabs}
\usepackage{longtable}
\usepackage{xcolor}

\title{Let's Do Turbulence Right: In Space and Time!}

\author{Maximilian Plattner$^{1}$\quad Fabian Paischer$^{1,2}$\quad Johannes Brandstetter$^{1,2}$\\
\normalfont $^{1}$ELLIS Unit, LIT AI Lab, Institute for Machine Learning, JKU Linz\\
\normalfont $^{2}$Mistral AI\\
\normalfont \href{mailto:plattner@ml.jku.at}{\texttt{plattner@ml.jku.at}}}

\iclrfinalcopy
\hypersetup{
  pdftitle={Let's Do Turbulence Right: In Space and Time!},
  pdfauthor={Maximilian Plattner, Fabian Paischer, Johannes Brandstetter}
}
\begin{document}

\maketitle
\vspace{-11pt}
\lhead{Preprint}

\begin{abstract}
The three-dimensional Navier–Stokes equations govern fluid turbulence, a chaotic and non-linear physical phenomenon characterized by complex eddies interacting across vast spatial and temporal scales. A numerically accurate modeling of this multi-scale energy cascade requires fully resolving both the fine spatial structures and the continuous temporal evolution of the flow field. In this paper, we leverage massive, time-resolved direct numerical simulations, which exceed 28 TB per run, to assess how numerical choices govern the resolved $(3+1)$-dimensional physics of turbulent flows.
Building on this analysis, we study how neural networks represent the recorded turbulent motion. Our results highlight that spatial and temporal assessments can disagree for both
numerical records and learned representations. We therefore call on the machine learning community to treat $(3+1)$-dimensional turbulence as a four-dimensional phenomenon.
\end{abstract} 

\begin{figure}[!htb]
\centering
\includegraphics[width=\linewidth]{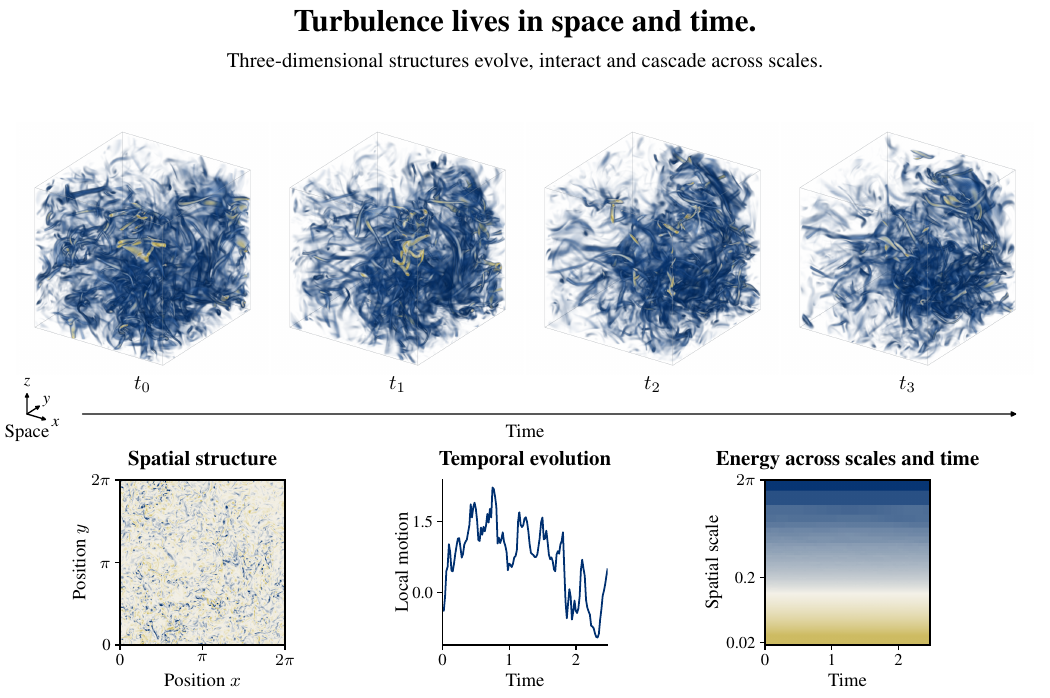}
\caption{The visualization shows \textbf{turbulent} structures from a single 28.9 TB time-resolved \textbf{direct numerical simulation}. The evolution of these structures (top) reveals coupled motion across space and time (bottom). How \textbf{neural representations} capture these \textbf{coupled spatio–temporal dynamics} remains an open question. To address this gap, we first use massive time-resolved simulation records to assess how numerical choices shape turbulent dynamics. We then examine how neural networks represent the recorded motion. Our analyses indicate that spatial and temporal assessments can disagree. \emph{We therefore call on the machine learning community to treat
$(3+1)$-dimensional turbulence as a four-dimensional phenomenon.}
}

\label{fig:turbulence-in-space-and-time}
\end{figure}

\clearpage


\section{Introduction}
\label{sec:introduction}

Fluid flows become \textbf{turbulent} when inertial effects outweigh viscous resistance, as measured by the Reynolds number. This creates a complex multiscale cascade where energy transfers from large vortices to small eddies before dissipating as heat. Turbulence is a ubiquitous phenomenon, governing transport in engineering flows, the atmosphere,
and the oceans~\citep{pope2000turbulent}. Its defining characteristics involve interacting structures across multiple scales in \textbf{space and time}, i.e., in $(3+1)$ dimensions (Figure~\ref{fig:turbulence-in-space-and-time}).

Resolving this multiscale motion through \textbf{direct numerical simulation}
(DNS) is computationally demanding.
For many industrial applications, simulations instead describe time-averaged
or spatially filtered flows, using closure models to account for unresolved
motion~\citep{reynolds1895dynamical,smagorinsky1963general}.
Much of the machine learning literature likewise focuses on selected aspects
of turbulence, including closures for reduced
simulations~\citep{ling2016reynolds}, two-dimensional
dynamics~\citep{kochkov2021machine}, and individual three-dimensional
snapshots~\citep{lienen2024zero,shenoy2026gaussian}. However, none of these settings address how \textbf{neural representations} capture the \textbf{coupled spatio-temporal dynamics} of genuine $(3+1)$-dimensional turbulence.

We therefore ask: \emph{How do numerical choices shape the coupled space–time dynamics resolved by DNS, and how do neural representations capture them?}

To address this question, we first assess how numerical modeling and representation shape the resulting dynamics of $(3+1)$-dimensional turbulent structures, drawing on massive, time-resolved direct numerical simulations, which exceed 28 TB per run. Building on this analysis, we study how neural networks represent the recorded turbulent motion.

We summarize our contributions as follows:

\begin{enumerate}
\item \textbf{Numerical study.} We characterize how resolution, filtering, solver
precision, and record quantization affect spatial and temporal behavior,
including numerical failures and storage requirements.
\item \textbf{Neural network study.} We compare joint space--time representations
with independently fitted snapshots, examining how data construction shapes
representation quality and when spatial and temporal assessments lead to
different conclusions.
\item \textbf{Data.} We prepare filtered DNS records at multiple coupled space–time bandwidths from our high-resolution, time-resolved 28.9 TB reference trajectory, with metadata documenting the source simulation and each record’s filtering, sampling, and storage.
\end{enumerate}

Overall, we show that spatial and temporal assessments can disagree for both
numerical records and learned representations.
The difficulty of representing DNS data therefore extends beyond the cost of
generating and storing it. \emph{We therefore call on the machine learning community to treat
$(3+1)$-dimensional turbulence as a four-dimensional phenomenon.}

\section{Characteristics of \texorpdfstring{$(3+1)$}{(3+1)}-Dimensional Turbulence}
\label{sec:turbulence-characteristics}

The three-dimensional Navier--Stokes equations describe fluid motion across different flow
regimes.
For an incompressible fluid, the velocity field $\uvel(\pos,t)$ obeys~\citep{batchelor1967introduction}
\begin{equation}
\label{eq:navier-stokes}
\partial_t\uvel+(\uvel\cdot\nabla)\uvel
=-\nabla\press+\visc\nabla^2\uvel+\force,
\qquad \nabla\cdot\uvel=0.
\end{equation}
In~\eqref{eq:navier-stokes}, the first relation balances fluid acceleration against pressure, viscous,
and applied forces; $\press$ is pressure per unit density, $\visc$ is kinematic
viscosity, and $\force$ is the applied force.
The second equation states that fluid motion preserves volume.
The nonlinear transport term $(\uvel\cdot\nabla)\uvel$ couples motion at
different scales.

\begin{wrapfigure}[29]{R}{0.50\textwidth}
\setlength{\abovecaptionskip}{6pt}
\centering
\includegraphics[width=\linewidth]{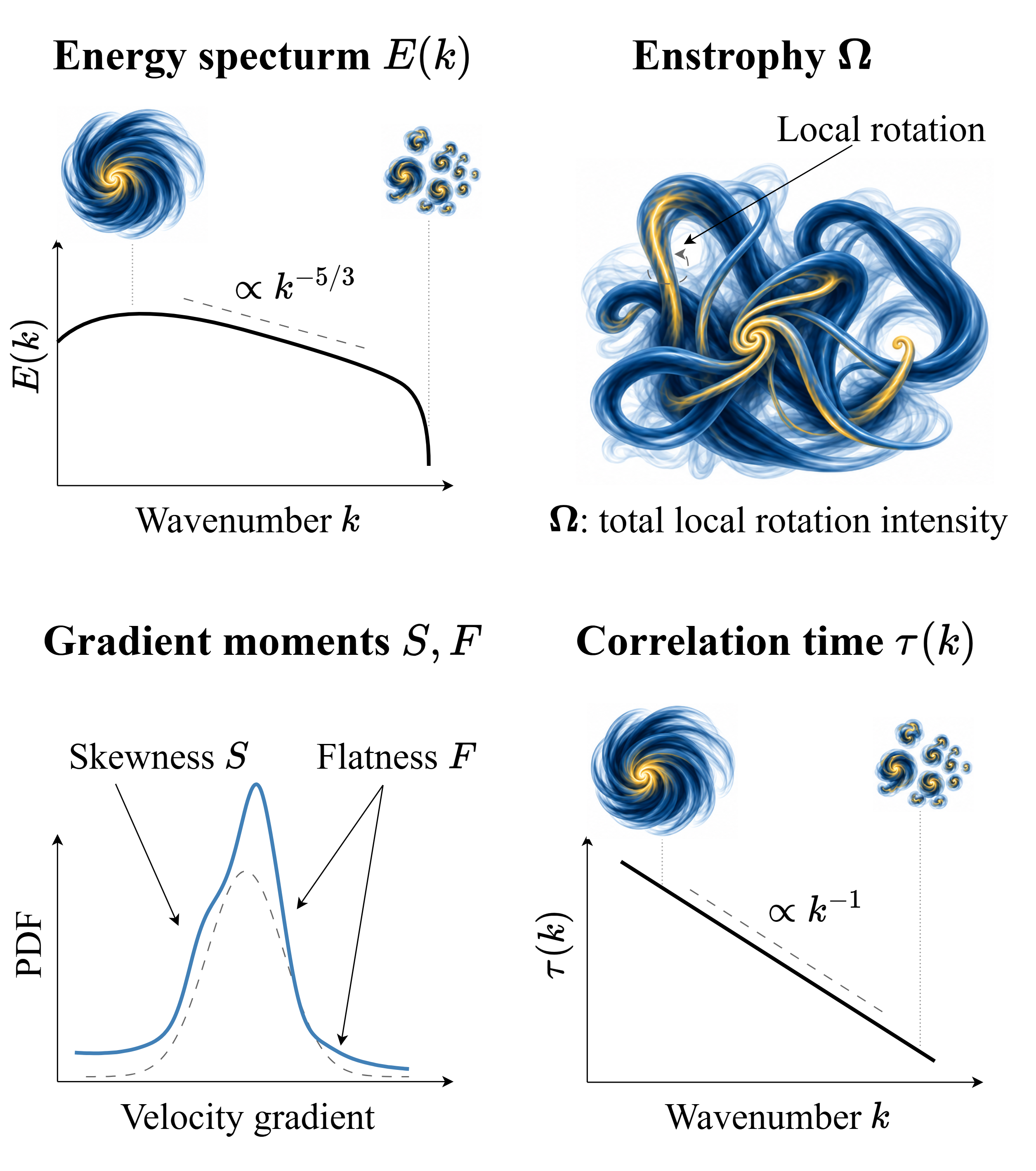}
\caption{\textbf{Turbulence diagnostics (schematic).} The energy spectrum shows how energy is distributed across spatial scales, with small scales carrying little energy but producing strong velocity gradients. These gradients generate local rotation, quantified by enstrophy. Their moments then reveal asymmetry and rare extreme events. Correlation time finally measures how quickly motion at each scale loses memory. Together, these diagnostics describe spatial and temporal dynamics.}
\label{fig:turbulence-statistics}
\end{wrapfigure}

This nonlinear transport competes with viscous smoothing.
Their relative importance is measured by the Reynolds number
$\Reyn=\urms\intscale/\visc$, where $\urms$ is the root mean square of one
velocity component and $\intscale$ is the integral length scale characterizing
the large motions.
When inertial effects dominate, nonlinear interactions can produce turbulent
motion across many scales in space and time. This motion can be characterized using statistical measures from turbulence theory~\citep{pope2000turbulent}. The following measures describe how energy is distributed across spatial scales, how strong and uneven velocity gradients are, and how motion at each scale decorrelates in time (Figure~\ref{fig:turbulence-statistics}).

\paragraph{Energy spectrum.}
The energy spectrum $E(k)$ describes
how kinetic energy is distributed across spatial scales.
It groups spatial Fourier modes by their wavenumber magnitude $k$, with
larger $k$ corresponding to smaller scales.
In developed turbulence, most energy lies at large scales.
An intermediate inertial range can exhibit $E(k)\propto k^{-5/3}$
~\citep{kolmogorov1941local,pope2000turbulent}.
At the smallest scales, viscosity causes a rapid spectral decay.

\paragraph{Enstrophy.}
Small-scale motion can carry little energy yet produce strong velocity gradients.
Vorticity $\boldsymbol\omega=\nabla\times\uvel$ describes local fluid rotation,
and enstrophy $\enst=\langle|\boldsymbol\omega|^2\rangle/2$ measures its mean
squared strength, where brackets denote spatial averages.
Spatial derivatives amplify contributions from large wavenumbers.
In turbulence, strong small-scale velocity gradients therefore contribute
substantially to enstrophy despite the low energy at these scales.

\paragraph{Gradient moments.}
Gradient moments describe how these velocity gradients are distributed. The \textit{skewness}
$S$ measures asymmetry between
positive and negative gradients.
In developed turbulence, $S$ is typically negative
~\citep{ishihara2009study}.
The \textit{flatness} $F$ measures the
relative weight of extreme gradients. In turbulence, $F$ typically exceeds the Gaussian value of $3$
~\citep{sreenivasan1997phenomenology}.

\paragraph{Correlation time.}
Temporal correlations describe how the spatial fluctuations characterized
above evolve at fixed coordinates.
For each wavenumber shell, the normalized correlation of velocity Fourier
coefficients at different time lags measures how quickly these fluctuations
lose correlation.
Its first crossing of $e^{-1}$ defines the correlation time $\tau_e(k)$.
In turbulence, large-scale motion carries smaller structures past fixed
positions, a process called sweeping.
When sweeping dominates, $\tau_e(k)\propto(\urms k)^{-1}$~\citep{tennekes1975eulerian,gorbunova2021spatio}.

\section{Direct Numerical Simulation (DNS)}
\label{sec:direct-numerical-simulation}

Numerical simulation provides access to the evolving velocity field, allowing
the statistics above to be studied under controlled physical conditions.
Direct numerical simulation (DNS) evolves the Navier--Stokes equations
(\ref{eq:navier-stokes}) with spatial grids and time steps fine enough to
resolve motion down to the scales at which viscosity dissipates
energy~\citep{moin1998direct}.
Saving states of this field produces a time-resolved velocity record
$\uvel^{\mathrm{rec}}$, which we also call a \emph{movie}: a sequence of
three-dimensional velocity fields (Figure~\ref{fig:turbulence-in-space-and-time}).
Its spatial and temporal content can be affected both by how the motion is
computed (\emph{numerical modeling}) and by how it is retained
(\emph{representation}).

\paragraph{Numerical modeling.}
Spatial \emph{resolution} determines which scales the solver can resolve.
For a cubic grid, let $N$ denote the number of points per spatial direction.
Arithmetic \emph{precision} determines the numerical values available during
time stepping: rounding enters the evolving state and can affect subsequent
motion.
We write $\uvel_{N,\sigma}$ for the record produced directly by the
simulation, where $\sigma$ denotes the solver's floating-point format.

\clearpage
\paragraph{Representation.}
A completed record can be transformed without rerunning the simulation.
\emph{Filtering}, denoted $\mathcal P_M$, selects the spatial and temporal
scales retained in that record.
It acts on space and time together, with temporal bandwidth chosen to
reflect how rapidly motion at the retained spatial scales evolves.
Here $M$ indexes the retained bandwidth and the resampled movie: $M$ points
per spatial direction and $M$ times give an $M^4$ array for each velocity
component.
\emph{Quantization}, denoted $\mathcal Q_q$, rounds recorded velocity values
to a floating-point format $q$.
Unlike rounding during evolution, these changes are not fed back into the
solver.
Thus $\uvel^{\mathrm{rec}}=\mathcal P_M\uvel_{N,\sigma}$ retains selected
scales of the computed trajectory, while
$\uvel^{\mathrm{rec}}=\mathcal Q_q\uvel_{N,\sigma}$ contains rounded velocity values.

\begin{wrapfigure}[43]{R}{0.43\textwidth}
\vspace{-\intextsep}
\vspace{-30pt}
\centering
\includegraphics[height=414.8pt,width=\linewidth,keepaspectratio]{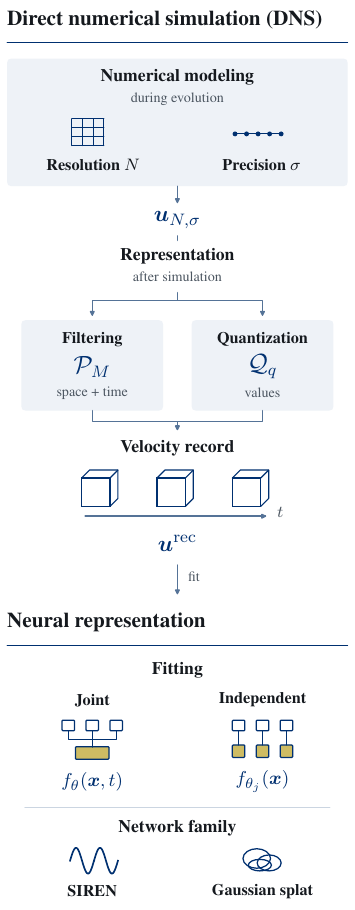}
\caption{\textbf{From DNS to neural representation.} Computation and recording determine the velocity samples supplied to joint or independent neural fitting.}
\label{fig:dns-neural-schematic}
\end{wrapfigure}

\section{Neural Representation}
\label{sec:neural-representations}

A DNS record supplies velocity samples at discrete positions and times.
A \emph{neural representation} approximates these samples with a continuous
function from coordinates to the three velocity components (Figure~\ref{fig:dns-neural-schematic}).
A standard way to fit its parameters is to minimize mean squared error
(MSE), the average squared difference between predicted and recorded
values~\citep{sitzmann2020implicit, shenoy2026gaussian}.
Fitting a supplied movie in this way represents recorded motion rather
than predicting its future.

\paragraph{Fitting procedures.}
A movie can be represented as one field evolving in time or as a sequence
of spatial fields.
In \emph{joint fitting}, one network
$f_\theta(\pos,t)\approx\uvel^{\mathrm{rec}}(\pos,t)$ takes position and time as inputs,
with parameters $\theta$ shared across the movie.
In \emph{independent fitting}, a separate network
$f_{\theta_j}(\pos)\approx\uvel^{\mathrm{rec}}(\pos,t_j)$ represents each recorded time
$t_j$.
Its parameters $\theta_j$ belong only to that snapshot.
Both fitting procedures approximate the same velocity samples, but only
joint fitting couples their representation through a shared function of
space and time.

\paragraph{Network families.}
The analysis of multiscale turbulence benefits from descriptions that
retain information about both scale and spatial
localization~\citep{farge1992wavelet}.
SIREN parameterizes a continuous field using a fully connected network
with sinusoidal activation functions~\citep{sitzmann2020implicit}.
Gaussian splatting uses spatially localized Gaussian primitives with
learned centers and shapes~\citep{kerbl2023gaussian}.
An additive velocity-field formulation sums such primitives with learned
vector-valued coefficients, giving a generalized Gaussian basis-function
network~\citep{poggio1990networks}.
We focus on SIRENs and additive Gaussian fields
because they distribute their representational capacity in fundamentally
different ways. SIREN uses a single coordinate network across the full domain,
whereas the Gaussian field constructs the velocity field from localized,
adaptive primitives.

\clearpage

\section{Experiments}
\label{sec:experiments}

In this paper, we want to understand how numerical choices shape the coupled space--time dynamics present in DNS records, and how neural representations capture the resulting dynamics. To that end, Section~\ref{sec:numerical-experiments} examines the effects of resolution, solver precision, filtering, and quantization on the recorded flow, and Section~\ref{sec:neural-representation-experiments} studies how these records are represented by joint space--time and independent snapshot-wise fits. Applying the same diagnostics from Section~\ref{sec:turbulence-characteristics} to both records and fits separates changes introduced during data construction from those introduced during neural fitting.


\begin{figure}[!ht]
\centering
\includegraphics[width=\linewidth]{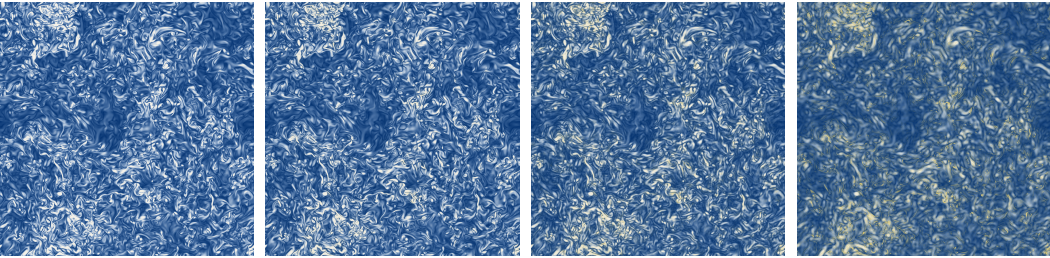}
\caption{\textbf{Space--time filtering.} From left to right, the DNS record
$\uvel_{1024,\mathrm{FP32}}$ is followed by its projections $\mathcal P_M$ at
$M=512,256,128$. Blue shows retained vorticity and yellow highlights the
removed part. Each panel shows the same time and plane $z=\pi$ on a common
intensity scale, with $0\leq x,y<2\pi$. Lighter shades reveal stronger
vorticity. We will release these filtered records and their construction metadata.}
\label{fig:filtering-structure}
\end{figure}

\subsection{Numerical modeling and representation}
\label{sec:numerical-experiments}

\begin{figure}[p]
\centering
\begin{minipage}{\linewidth}
\includegraphics[width=\linewidth]{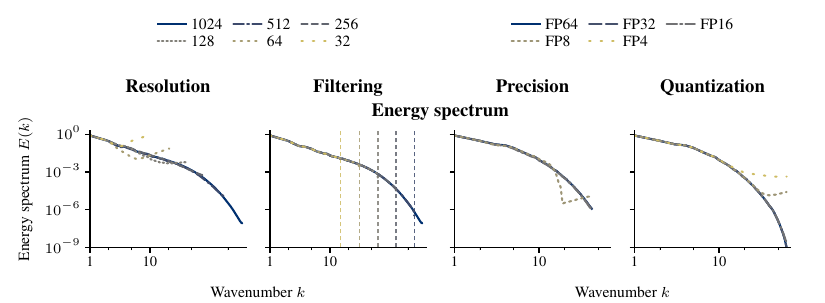}
\par\smallskip
{\normalsize\noindent\captiona\ The \textbf{energy spectrum} changes with \emph{resolution}:
coarse grids accumulate energy near their cutoff, while refinement extends the
resolved range. \emph{Filtering} removes fine scales but preserves the retained
spectrum. \emph{Precision} has little effect from FP64 to FP16, whereas FP8
distorts the small-scale decay. \emph{Quantization} adds small-scale energy as
the record format becomes coarser.\par}
\end{minipage}
\par\smallskip
\begin{minipage}{\linewidth}
\includegraphics[width=\linewidth]{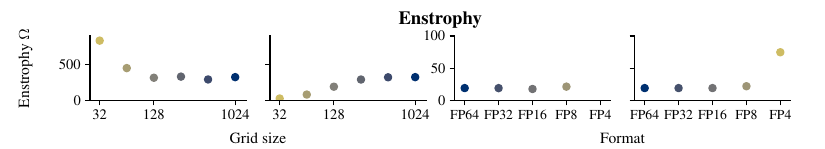}
\par\smallskip
{\normalsize\noindent\captionb\ \textbf{Enstrophy} increases at coarse \emph{resolution} because
energy accumulates near the grid cutoff. \emph{Filtering} removes substantial
gradient strength despite retaining most energy. \emph{Precision} changes
enstrophy only modestly through FP8. Coarse \emph{quantization} instead raises
enstrophy through numerical distortion.\par}
\end{minipage}
\par\smallskip
\begin{minipage}{\linewidth}
\includegraphics[width=\linewidth]{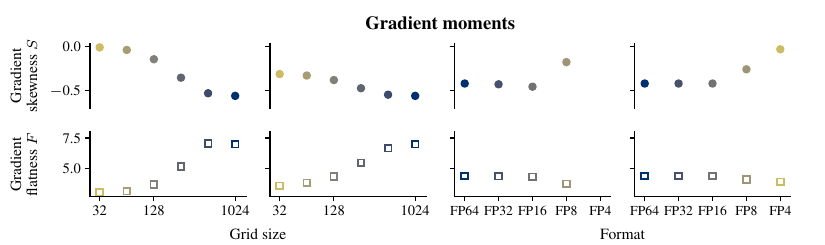}
\par\smallskip
{\normalsize\noindent\captionc\ The \textbf{gradient moments} change distinctly
across the four numerical choices. Increasing \emph{resolution} restores more
negative gradient skewness and heavier tails. Stronger \emph{filtering}
weakens both signatures. \emph{Precision} preserves them from FP64 to FP16,
whereas FP8 weakens them. Coarser \emph{quantization} likewise suppresses
gradient asymmetry and heavy tails.\par}
\end{minipage}
\par\smallskip
\begin{minipage}{\linewidth}
\includegraphics[width=\linewidth]{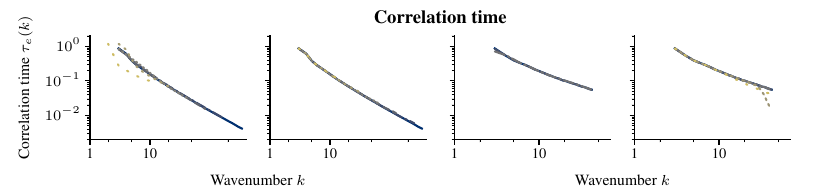}
\par\smallskip
{\normalsize\noindent\captiond\ \textbf{Correlation times} vary with \emph{resolution} even at
shared scales. \emph{Filtering} largely preserves the dynamics of the scales it
retains. For \emph{precision}, FP64--FP16 agree, FP8 is frozen, and FP4 produces
no evolved record. Coarse \emph{quantization} shortens small-scale
correlations.\par}
\end{minipage}
\caption{Numerical modeling and representation of the recorded flow.}
\label{fig:numerical-effects-on-turbulence}
\end{figure}

We write $\uvel_{N,\sigma}$ for the record produced directly by a simulation
with $N$ points per spatial direction and solver format $\sigma$.
The record analyzed is denoted by $\uvel^{\mathrm{rec}}$.
The operator $\mathcal P_M$ retains bandwidth $M$, while $\mathcal Q_q$ rounds
recorded values to format $q$.

\paragraph{Numerical modeling.}
For \emph{resolution}, we compare independent simulations
$\uvel^{\mathrm{rec}}=\uvel_{N,\mathrm{FP32}}$ at
$N=32,64,128,256,512,1024$.
We fix viscosity $\visc$ to test coarser grids on the high-Reynolds-number flow
resolved at $N=1024$.
For \emph{precision}, we use the higher-viscosity, spatially resolved
$N=128$ reference and compare
$\uvel^{\mathrm{rec}}=\uvel_{128,\sigma}$ with
$\sigma\in\{\mathrm{FP64},\mathrm{FP32},\mathrm{FP16},\mathrm{FP8},\mathrm{FP4}\}$.
Within each comparison, the initial condition, forcing, and recording schedule are fixed, as are the
nominal viscosity and time step.
FP64 uses native arithmetic and lower formats emulate the same operations with
nearest-value rounding after each arithmetic result.

\paragraph{Representation.}
For \emph{filtering}, we compare
$\uvel^{\mathrm{rec}}=\mathcal P_M\uvel_{1024,\mathrm{FP32}}$ at
$M=32,64,128,256,512$.
For \emph{quantization}, we compare
$\uvel^{\mathrm{rec}}=\mathcal Q_q\uvel_{128,\mathrm{FP64}}$, using the same
five formats for $q$ as for solver precision.
Each uses the corresponding nearest-value rounding rule on the completed
FP64 trajectory.

\paragraph{Effects on the recorded motion.}

Figure~\ref{fig:numerical-effects-on-turbulence} shows how the four numerical and representational choices affect both the spatial structure and the temporal evolution of the recorded flow (Appendix~\ref{app:numerical-outcomes}, Table~\ref{tab:numerical-all-records}). The \emph{energy spectrum} shows how energy is distributed across spatial scales, while \emph{enstrophy} captures the strength of the corresponding velocity gradients. Looking at the two together reveals an important difference: filtering removes small-scale gradient content, whereas insufficient resolution and coarse quantization can instead introduce artificial gradient strength. The \emph{gradient moments} provide a more detailed view of these gradients, showing that increased enstrophy does not necessarily recover the asymmetry and heavy tails associated with developed turbulence. The \emph{correlation times} then show how these structures evolve over time. Filtering largely preserves the dynamics of the scales that remain, while insufficient resolution and coarse quantization can alter their evolution. At very low solver precision, the dynamics can even become effectively frozen (Appendix~\ref{app:numerical-modeling-representation}, Figure~\ref{fig:numerical-joint-spectrum}). Thus, a record may look reasonable according to a spatial diagnostic while still failing to preserve the underlying turbulent motion over time.

\clearpage
\subsection{Neural representations}
\label{sec:neural-representation-experiments}

\begin{figure}[p]
\centering
\centerline{\includegraphics[width=\linewidth]{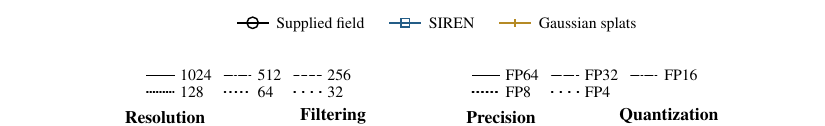}}
\par\smallskip
\begin{minipage}{\linewidth}
\centerline{\includegraphics[width=\linewidth]{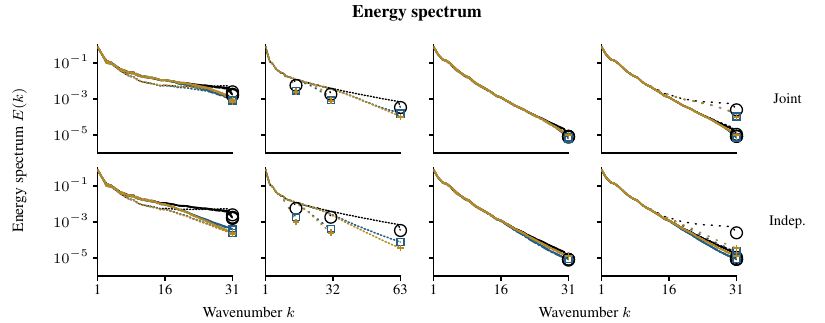}}
\par\smallskip
{\normalsize\noindent\captiona\ The \textbf{energy spectrum} retains the energy-containing
scales under both \emph{joint} and \emph{independent fitting}. Small-scale energy
is generally underestimated, especially by independent fits. Across
\emph{resolution} and \emph{filtering}, this removes supplied structure.
For coarse \emph{quantization}, it attenuates numerical excess.\par}
\end{minipage}
\par\medskip
\begin{minipage}{\linewidth}
\centerline{\includegraphics[width=\linewidth]{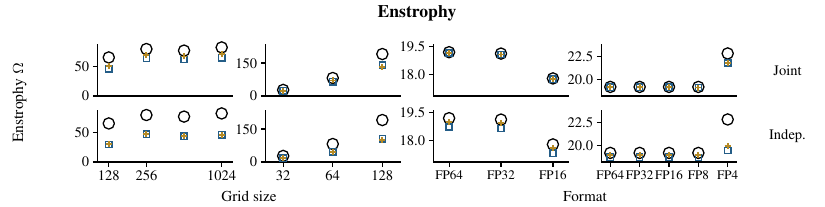}}
\par\smallskip
{\normalsize\noindent\captionb\ \textbf{Enstrophy} is generally lower under
\emph{independent fitting} than under \emph{joint fitting}. Deficits occur across
source \emph{resolutions} and grow with wider \emph{filtering} bandwidths.
On the FP4-quantized record, this reduction instead brings enstrophy closer
to the unquantized value.\par}
\end{minipage}
\par\medskip
\begin{minipage}{\linewidth}
\centerline{\includegraphics[width=\linewidth]{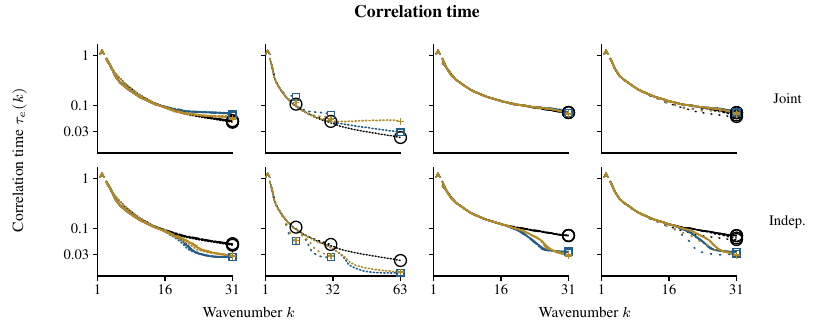}}
\par\smallskip
{\normalsize\noindent\captionc\ \textbf{Correlation times} expose a temporal trade-off.
\emph{Independent fits} generally decorrelate too rapidly at small scales,
whereas \emph{joint fits} can remain correlated too long. This tendency spans
\emph{resolution}, \emph{filtering}, \emph{precision}, and \emph{quantization}.\par}
\end{minipage}
\caption{Joint and independent neural representations of the recorded flow.}
\label{fig:joint-versus-independent-neural-representations}
\end{figure}

The numerical study establishes what motion is available in each record.
We now ask how neural representations capture that motion.
We compare joint fitting, which shares one model across the movie, with
independent fitting, which assigns a separate model to each snapshot.
Both use SIREN and Gaussian fields (Section~\ref{sec:neural-representations}),
with velocity MSE, comparable total parameter counts, and a common stopping rule.

\paragraph{Learning targets.}
The supplied records follow the four comparisons in
Section~\ref{sec:numerical-experiments}. For resolution, precision, and
quantization, we apply the same space--time projection to obtain a common
observation bandwidth:
resolution, $\uvel^{\mathrm{rec}}=\mathcal P_{64}\uvel_{N,\mathrm{FP32}}$;
precision, $\uvel^{\mathrm{rec}}=\mathcal P_{64}\uvel_{128,\sigma}$;
quantization, $\uvel^{\mathrm{rec}}=\mathcal P_{64}\mathcal Q_q\uvel_{128,\mathrm{FP64}}$.
For resolution, $N$ ranges over $128,256,512,1024$; for precision,
$\sigma$ ranges over the evolving FP64, FP32, and FP16 records.
Quantization uses all five formats for $q$, with rounding before
projection. The filtering comparison varies the observation bandwidth itself:
$\uvel^{\mathrm{rec}}=\mathcal P_M\uvel_{1024,\mathrm{FP32}}$ for $M=32,64,128$. Each fit is assessed against its supplied movie.

\paragraph{Effects of neural fitting.}

Figure~\ref{fig:joint-versus-independent-neural-representations} compares how well the joint and independent fits reproduce both the spatial structure and the temporal evolution of the supplied records (Appendix~\ref{app:neural-absolute-diagnostics}, Figure~\ref{fig:neural-representation-comparison}, Tables~\ref{tab:neural-diagnostics-resolution}--\ref{tab:neural-diagnostics-quantization}). The \emph{energy spectrum} shows which scales are lost or weakened, while \emph{enstrophy} captures the change in gradient strength. Whether this attenuation is desirable depends on the record: fitting can remove genuine turbulent structure or suppress spurious content introduced by quantization. The \emph{gradient moments} provide a complementary view. Similar enstrophy does not necessarily mean that the underlying gradients are distributed in the same way. SIREN and Gaussian fits differ in skewness and flatness even after matching gradient power across scales (Appendix~\ref{app:neural-controls}, Figure~\ref{fig:gradient-organization}). The \emph{correlation times} then reveal differences invisible to these spatial diagnostics. Joint fits generally retain fluctuations for longer, whereas independent fits decorrelate more quickly, with neither consistently matching the supplied dynamics better (Appendix~\ref{app:neural-evaluation}, Table~\ref{tab:neural-seed-sensitivity}; Appendix~\ref{app:neural-controls}, Figure~\ref{fig:neural-joint-spectrum}). Taken together, these results show that spatial and temporal fidelity are not interchangeable: comparable frame accuracy can still yield different accuracy of the recorded changes (Appendix~\ref{app:neural-controls}, Table~\ref{tab:matched-trained-error}).

\section{Discussion and key takeaways}

Our results show that spatial fidelity and temporal fidelity are distinct
properties of a turbulent representation. This distinction already appears
before learning. At fixed physical parameters, insufficient resolution can
create excess small-scale energy, distort gradients, and alter correlation
times. Space-time filtering instead removes fine scales while largely preserving the
temporal behavior of those retained. Low solver precision can preserve a
structured field while arresting its evolution, whereas post-hoc quantization
can add apparent small-scale content while shortening correlations. Thus,
spatial detail alone does not determine whether a record contains faithful
turbulent motion.

Neural representations inherit this ambiguity. Both network families attenuate small-scale content, but whether this is beneficial depends on its origin: fitting can remove genuine turbulent gradients from a resolved record, yet suppress numerical distortion in a strongly quantized one. This parallels the mitigation of numerical artifacts observed in neural PDE emulators \citep{koehler2025neural}. Likewise, representations with similar enstrophy can have different gradient statistics, and similar spatial accuracy can coexist with different temporal correlations. The simulation, recording procedure, and learned representation should therefore be viewed as a single information pipeline: each stage determines which spatial and temporal information remains available to the next.

In practice, our results suggest that the architecture and temporal parameter sharing of neural representations should be matched to the gradients and velocity changes relevant to the application. For example, joint Gaussian fits better preserve the record's small-scale temporal correlations than independent fits. These choices become increasingly important as neural representations are used to encode turbulent flows for downstream applications \citep{du2024conditional,guo2025conditional,lu2026latent}. Furthermore, practitioners should account for how numerical and recording choices affect the spatial and temporal information these representations preserve, as this information can shape the turbulent dynamics downstream models learn to reproduce. For instance, independent SIREN fits show a smoothing bias that counteracts excess enstrophy caused by quantization.

\clearpage
\section{Related Work}
\label{sec:related-work}
\paragraph{Numerical modeling of turbulence.}
DNS resolves three-dimensional turbulent motion in space and time, down to
the scales at which viscosity dissipates energy
\citep{moin1998direct,ishihara2009study}.
Its computational cost motivates a hierarchy of reduced descriptions.
As the Reynolds number increases, the separation between large and small
scales grows, requiring finer grids and shorter time steps.
For many industrial applications, resolving this full range is impractical.
Reynolds-averaged Navier--Stokes (RANS) models instead describe the mean
flow~\citep{reynolds1895dynamical}, while large-eddy simulation (LES)
resolves larger motions and models unresolved
scales~\citep{smagorinsky1963general,germano1991dynamic}.
Both require closures because averaging or filtering the nonlinear term
in~\eqref{eq:navier-stokes} introduces stresses that cannot be determined
from the retained velocity alone.
For velocity components $u_i,u_j$, the difference
$\overline{u_i u_j}-\bar u_i\bar u_j$ expresses this unresolved contribution,
where the bar denotes the chosen average or spatial filter.
The closure supplies its effect on the retained motion.
A different simplification restricts motion to two spatial dimensions.
This reduces the computational burden but also changes the cascade.
Without vortex stretching, the inviscid, unforced dynamics conserve both
energy and enstrophy. In the classical two-dimensional cascade, energy
therefore transfers toward larger scales, while enstrophy transfers toward
smaller scales~\citep{boffetta2012two}.
Two-dimensional turbulence thus offers a simpler setting for studying
multiscale motion, with different energy-transfer physics from the
three-dimensional flow considered here.

\paragraph{Machine learning for turbulence.}
Machine learning builds on these reduced approaches, using simulated flows
for training and evaluation
\citep{brunton2020machine,duraisamy2019turbulence}.
What a model learns therefore depends in part on how those flows were
computed and recorded.
Applications include learned closures for unresolved
stresses~\citep{ling2016reynolds} and learned numerical schemes for
coarse-grid evolution in two-dimensional turbulence~\citep{kochkov2021machine}.
Super-resolution methods recover spatial detail from coarser observations
\citep{fukami2019super,fukami2024single}, and generative approaches synthesize
three-dimensional turbulent states~\citep{lienen2024zero}.
Much of this work addresses selected aspects of the full $(3+1)$-dimensional
motion resolved by DNS.

\paragraph{Neural representations for turbulence.}
More recently, methods have begun to generate
three-dimensional flow sequences
\citep{du2024conditional}.
Neural representations offer one way to encode such sequences. For
instance, CoNFiLD encodes flow fields with a conditional neural
field~\citep{du2024conditional}.
Related neural representations for turbulence have also been explored
\citep{pan2023neural,guo2025conditional,shenoy2026gaussian}.
What these representations preserve, however, depends on their inductive biases: small velocity errors, for example, can coexist with substantial enstrophy loss~\citep{shenoy2026gaussian}. 
These biases act on flows that are themselves shaped by numerical and recording choices, raising the question of how such choices interact with the way space and time are represented and fitted.

\section{Conclusion}
\label{sec:conclusion}

In this work, we examine how numerical choices shape turbulent motion in space
and time, and how neural networks represent the resulting records.
This matters because what these representations preserve determines which
turbulent dynamics downstream models can learn
\citep{du2024conditional,guo2025conditional,lu2026latent}.
To the best of our knowledge, this is the first study to trace how choices made
during simulation and recording propagate into neural representations of the
full $(3+1)$-dimensional motion resolved by DNS.
Using massive time-resolved DNS records, we vary resolution, solver precision,
filtering, and quantization, and compare joint space--time fitting with
independent snapshot fitting.
We find that spatial and temporal assessments can disagree for both
numerical records and learned representations.
The difficulty of representing DNS data therefore extends beyond the cost of
generating and storing it.
\emph{We therefore call on the machine learning community to treat
$(3+1)$-dimensional turbulence as a four-dimensional phenomenon.}

\label{sec:main-text-end}
\clearpage


\subsection*{AI use statement}

In this work, we used generative AI tools for two tasks requiring disclosure: generating synthetic datasets and implementing methods. We used Claude to help set up the code for the simulation environment used to generate the data discussed in Section \ref{sec:numerical-experiments}. One author reviewed and verified both this code and the resulting simulation record. We also used Codex to implement the neural representations (SIREN and Gaussian splats) and the supporting evaluation code used in Section \ref{sec:neural-representation-experiments}. One author reviewed, verified, and tested this code.

We did not use generative AI tools to develop the problem setting or framing; to help develop theoretical models or conceptual frameworks; to formulate mathematical claims; to propose or refine hypotheses; to design or provide feedback on research methodology or experiments; to clean or reformat datasets; or to interpret results. Providing critical ingredients for proving mathematical claims, assisting in the writing of proofs, translation, and qualitative or thematic data analysis were not applicable to this work.

Additionally, we used generative AI tools to identify relevant literature, improve the manuscript’s readability, and make cosmetic improvements to figures. We did not use these tools for other tasks for which disclosure is recommended, such as drafting text, suggesting hyperparameters, identifying research gaps, formatting references, or proposing a title.

We have reviewed all AI-assisted work. In addition to the code checks described above, all suggested papers were read by the authors before being cited, and text edits were checked to ensure that the technical content and claims remained unchanged. Figure edits were limited to styling. We take responsibility for the final content of this work, including text, claims or artifacts produced with the aid of generative AI.

\subsection*{Reproducibility statement}
Appendix~\ref{app:numerical-modeling-representation} specifies the physical
setup, solver, initial conditions, time stepping, arithmetic formats,
recording schedules, filtering, quantization, and diagnostic estimators.
Appendix~\ref{app:neural-experiments} specifies the learning targets,
architectures, parameter budgets, normalization, sampling, optimizer,
stopping criterion, initialization, and evaluation procedure.

\clearpage


\clearpage
\appendix

\section{Characteristics of \texorpdfstring{$(3+1)$}{(3+1)}-Dimensional Turbulence}
\label{app:turbulence-characteristics}

Turbulence is a state of fluid motion with fluctuations over many spatial
scales and time scales. This motion is described by the velocity field
$\uvel(\pos,t)$. This field has three components at each position
$\pos=(x_1,x_2,x_3)$ and time $t$. The four dimensions therefore describe
three spatial coordinates and one temporal coordinate. The equations and statistical quantities
that follow describe its spatial and temporal structure. We refer to
\citet{pope2000turbulent} for a fuller treatment of turbulence theory.

For a fluid of constant density and kinematic viscosity $\visc$, the
incompressible Navier--Stokes equations~(\ref{eq:navier-stokes}) read
\begin{equation}
 \underbrace{\partial_t\uvel+(\uvel\cdot\nabla)\uvel}_{\text{material acceleration}}
 =-\nabla\press+\visc\nabla^2\uvel+\force,
 \qquad \nabla\cdot\uvel=0.
 \label{eq:app-ns}
\end{equation}
Here $\press$ is pressure divided by density and $\force$ is force per unit
mass. The time derivative measures change at a fixed position. The transport
term accounts for a fluid parcel moving through a spatially varying field.
Their sum is the acceleration experienced by that parcel. Nonlinear
transport couples spatial scales, while viscosity smooths velocity gradients.
Thus the equation connects \textbf{spatial variation and temporal evolution}.

\paragraph{Incompressibility.}
The condition $\nabla\cdot\uvel=\sum_i\partial u_i/\partial x_i=0$ means
that an infinitesimal fluid volume neither expands nor contracts. Pressure enforces this constraint: taking
the divergence of~(\ref{eq:app-ns}) gives
\begin{equation}
 \nabla^2\press
 =-\sum_{i,j}(\partial_i u_j)(\partial_j u_i)+\nabla\cdot\force,
 \qquad \partial_i=\frac{\partial}{\partial x_i}.
 \label{eq:app-pressure}
\end{equation}
The root-mean-square divergence
$\langle(\nabla\cdot\uvel)^2\rangle^{1/2}$ quantifies departures from this
constraint. It vanishes for an incompressible field. Here
$\langle a\rangle=|V|^{-1}\int_V a\,d\pos$ denotes a spatial average over
the domain $V$. An overbar denotes a time average.

\paragraph{Statistical setting and energy balance.}
\emph{Homogeneity} means that statistics do not depend on position, and
\emph{isotropy} that they do not depend on orientation.  \emph{Stationarity} means that
statistics do not change with the time origin.
Consider a zero-mean velocity field in a periodic domain. Its kinetic energy
per unit mass is
$K=\langle|\uvel|^2\rangle/2$. Multiplying~(\ref{eq:app-ns}) by velocity
and averaging gives
\begin{equation}
 \frac{dK}{dt}=\langle\uvel\cdot\force\rangle-\diss,
 \qquad
 \diss=2\visc\left\langle\sum_{i,j}s_{ij}^2\right\rangle,
 \qquad s_{ij}=\tfrac12(\partial_j u_i+\partial_i u_j).
 \label{eq:app-energy-balance}
\end{equation}
The symmetric tensor $s_{ij}$ measures local deformation. The quantity
$\diss$ is the rate of viscous energy loss. Nonlinear transport redistributes
energy without changing its total in this domain. Forcing supplies energy.
In a statistically stationary flow, its mean input balances mean dissipation.

\paragraph{Reynolds number and the range of scales.}
Let $\urms=(2\overline K/3)^{1/2}$ be the root-mean-square velocity of one
component in isotropic flow, and $\intscale$ the integral length scale,
which characterizes the extent of large-scale velocity correlations. Then
\begin{equation}
 \Reyn=\frac{\urms\intscale}{\visc},\qquad
 T_L=\frac{\intscale}{\urms},\qquad
 \eta=\left(\frac{\visc^3}{\overline\diss}\right)^{1/4},\qquad
 \tau_\eta=\left(\frac{\visc}{\overline\diss}\right)^{1/2}.
 \label{eq:app-turbulent-scales}
\end{equation}
The Reynolds number compares inertial transport with viscous smoothing.
The turnover time $T_L$ describes the large motions. The Kolmogorov scales
$\eta$ and $\tau_\eta$ characterize viscous dissipation. Large Reynolds
numbers permit a broad separation between them. This separation allows
motions at many scales to coexist. An alternative measure is the
\emph{Taylor-scale Reynolds number}
$\Relam=\urms\lambda/\visc$, where
$\lambda^2=15\visc(\urms)^2/\overline\diss$ for isotropic incompressible
flow. Both Reynolds numbers express the relative importance of inertia and
viscosity, but they use different characteristic length scales.

\clearpage
\paragraph{Energy spectrum.}
A spatial Fourier expansion writes
$\uvel(\pos,t)=\sum_{\wave}\uhat(\wave,t)e^{i\wave\cdot\pos}$.
A mode of magnitude $k=|\wave|$ has wavelength $2\pi/k$.
Grouping modes in radial shells gives
\begin{equation}
 E(k,t)=\frac12\sum_{\wave\in\mathcal S_k}|\uhat(\wave,t)|^2,
 \qquad K(t)=\sum_k E(k,t),
 \label{eq:app-shell-energy}
\end{equation}
where $\mathcal S_k$ is the shell labeled $k$ and the Fourier normalization
preserves spatial mean-square velocity. The continuum spectrum instead
satisfies $K=\int_0^\infty E(k)\,dk$.
In developed three-dimensional turbulence, energy is supplied at large
scales and transferred on average toward smaller scales. Where forcing
and viscosity have little direct effect, Kolmogorov theory predicts
$E(k)\simeq C_K\overline\diss^{\,2/3}k^{-5/3}$, with dimensionless constant
$C_K$~\citep{kolmogorov1941local,pope2000turbulent}.
This \emph{inertial range} appears only when the large and dissipative scales
are sufficiently separated.

\paragraph{Enstrophy.}
Vorticity $\vort=\nabla\times\uvel$ measures local rotation, and
$\enst=\langle|\vort|^2\rangle/2$ is its enstrophy. For a smooth,
periodic, divergence-free field,
\begin{equation}
 \enst=\frac12\sum_{\wave}|\wave|^2|\uhat(\wave)|^2,
 \qquad \diss=2\visc\enst.
 \label{eq:app-enstrophy}
\end{equation}
The factor $|\wave|^2$ makes fine scales important even when they carry
little energy. In continuum notation, $\enst=\int k^2E(k)\,dk$.
The exact Fourier sum weights each mode by its own wavenumber.
For a general periodic velocity field,
$\langle|\nabla\uvel|^2\rangle
=2\enst+\langle(\nabla\cdot\uvel)^2\rangle$.
The relation $\diss=2\visc\enst$ therefore relies on incompressibility.

\paragraph{Gradient moments.}
The longitudinal derivatives $g_i=\partial_i u_i$ have zero spatial mean
in a periodic domain. Their normalized third and fourth moments are
\begin{equation}
 S_i=\frac{\langle g_i^3\rangle}{\langle g_i^2\rangle^{3/2}},
 \qquad F_i=\frac{\langle g_i^4\rangle}{\langle g_i^2\rangle^2},
 \qquad S=\tfrac13\sum_i S_i,\quad F=\tfrac13\sum_i F_i.
 \label{eq:app-gradient-moments}
\end{equation}
These moments are first defined through spatial averages. Their stationary
values can then be averaged over time. Negative \emph{skewness} indicates a
greater contribution from negative gradients to the third moment.
\emph{Flatness} above the Gaussian value $3$ indicates heavier tails.
Developed turbulence typically has negative skewness, often near $-0.5$,
and flatness above $3$. The values
depend on Reynolds number and retained scales~\citep{ishihara2009study,sreenivasan1997phenomenology}.

\paragraph{Correlation time.}
The preceding quantities describe the spatial field at one time. Temporal
correlation describes how long each spatial scale retains its state. Define
the shell correlation
\begin{equation}
 R(k,\tau)=\operatorname{Re}\sum_{\wave\in\mathcal S_k}
 \overline{\uhat(\wave,t)^*\cdot\uhat(\wave,t+\tau)},
 \qquad C(k,\tau)=\frac{R(k,\tau)}{R(k,0)}.
 \label{eq:app-shell-correlation}
\end{equation}
The star denotes complex conjugation. The overbar averages over the time
origin $t$. For nonzero shell energy, $C(k,0)=1$. The first crossing of
$e^{-1}$ defines the correlation time $\tau_e(k)$. A frozen nonzero field
has $C(k,\tau)=1$ at every lag. An empty shell has no normalized correlation.

These are Eulerian correlations: they describe change at fixed
coordinates, not the lifetime of a moving vortex. Large motions can sweep
small structures past those coordinates without destroying them.
When sweeping dominates, the characteristic time scales as
$\tau_e(k)\propto(\urms k)^{-1}$~\citep{tennekes1975eulerian,gorbunova2021spatio}.

\clearpage
\section{Direct Numerical Simulation (DNS)}
\label{app:dns-framework}

Appendix~\ref{app:turbulence-characteristics} describes turbulent motion as
a velocity field evolving across spatial and temporal scales.
Direct numerical simulation (DNS) computes this field from the
Navier--Stokes equations without a closure for unresolved motion
and resolves the scales of viscous dissipation~\citep{moin1998direct}.
Section~\ref{sec:direct-numerical-simulation} distinguishes the \emph{numerical modeling}
that computes this trajectory from the \emph{representation} that retains it
as a record.

\subsection{Numerical modeling}
\label{app:dns-modeling}

\paragraph{From the equations to a solver.}
The physical problem specifies the domain, boundary conditions, viscosity
$\visc$, forcing $\force$, and initial velocity $\uvel(\pos,0)$.
Consider a periodic cube with side length $\domside$.
Opposite faces connect, which permits a Fourier representation of the field.
With \textbf{spatial resolution $N$}, each direction contains $N$ grid
points and the grid spacing is $\dx=\domside/N$.
The truncated field $\uvel_N$ uses a finite set of Fourier modes $\mathcal K_N$.
Projecting~(\ref{eq:app-ns}) onto these modes gives, for $\wave\ne0$,
\begin{equation}
 \frac{d\uhat_N(\wave,t)}{dt}
 =-\mathsf P_{\wave}\widehat{(\uvel_N\cdot\nabla)\uvel_N}(\wave,t)
  -\visc|\wave|^2\uhat_N(\wave,t)
  +\mathsf P_{\wave}\widehat{\force}(\wave,t),
 \qquad
 \mathsf P_{\wave}=I-\frac{\wave\wave^{\mathsf T}}{|\wave|^2}.
 \label{eq:app-spectral-evolution}
\end{equation}
Here $I$ is the identity matrix.
The projection $\mathsf P_{\wave}$ enforces
$\wave\cdot\uhat_N=0$ and thus eliminates pressure from the evolution equation.
The zero mode describes the spatial mean and evolves with the mean forcing.

Equation~(\ref{eq:app-spectral-evolution}) defines a finite system of
ordinary differential equations.
A \emph{pseudo-spectral} method evaluates derivatives in Fourier space and
nonlinear products in physical space.
These products generate modes outside $\mathcal K_N$.
On a finite grid, unresolved modes can reappear at false lower wavenumbers.
This effect is \emph{aliasing}.
Dealiasing prevents it through padding or a restriction of the retained modes.
The dealiasing rule determines $\kmaxi$, the largest retained shell
wavenumber.
Thus $N$ alone does not specify the resolved band.

\paragraph{Time stepping and arithmetic.}
A time integrator advances the retained coefficients over intervals
$\dtsolv$.
Let $U^n$ collect the numerical state at solver time $n\dtsolv$.
Runge--Kutta methods approximate evolution through intermediate stages.
An integrating factor can apply the viscous decay
$e^{-\visc|\wave|^2\dtsolv}$ directly.

The computer evaluates each stage with finite arithmetic.
A floating-point format provides a finite set of real numbers
$\mathbb F_\sigma$.
Away from zero, a binary number has the form $\pm m\,2^e$, with $1\le m<2$.
The significant digits in $m$ determine the spacing of nearby values.
The allowed exponents determine the range of magnitudes~\citep{goldberg1991floating}.
Let $r_\sigma(a)$ round a real value $a$ to the nearest available value.
Ties, very small values, and values beyond the finite range require specified rules.

With \textbf{arithmetic precision $\sigma$}, one solver step has the form
\begin{equation}
 U^{n+1}=\Phi_{N,\sigma}^{\dtsolv}(U^n).
 \label{eq:app-solver-step}
\end{equation}
The map $\Phi_{N,\sigma}^{\dtsolv}$ includes the nonlinear term, forcing,
viscous evolution, and the incompressibility projection.
Rounding can enter its intermediate stages, transforms, constants, and state.
Each rounded result can affect every later step.
Rounding only the completed update defines a different map.
The meaning of $\sigma$ therefore includes the format, rounding rules, and
the operations to which those rules apply.

\paragraph{From solver states to a record.}
Repeated application of~(\ref{eq:app-solver-step}) produces the computed
trajectory.
Saving selected states produces the record $\uvel_{N,\sigma}$.
For $J$ equally spaced snapshots, write
$t_j=t_0+j\dtsnap$, with $j=0,\ldots,J-1$.
The snapshot interval $\dtsnap$ differs from the solver step $\dtsolv$.
The observation span is $(J-1)\dtsnap$.
Reproducing the trajectory requires the initial condition, forcing,
time integrator, dealiasing rule, and arithmetic rules.
Reproducing the record also requires the discarded transient, snapshot times,
stored fields, and encoding.

The ratios $\kmaxi\eta$ and $\dtsolv/\tau_\eta$ compare the solver's
resolution with the dissipative scales defined in
Appendix~\ref{app:turbulence-characteristics}.
The advective Courant number
$\dtsolv\max_{\pos}\sum_i|u_i|/\dx$ also constrains time-step stability.
Snapshot spacing is described by $\dtsnap/\tau_\eta$ and, at each shell,
$\dtsnap/\tau_e(k)$.
The observation span in units of $T_L$ describes coverage of large-scale motion.
Measured $\urms$, $\overline\diss$, $L$, and $\Relam$ characterize the
resulting flow.

\clearpage
\subsection{Representation}
\label{app:dns-representation}

A record retains a finite description of the computed trajectory.
Its representation determines which scales and velocity values remain.
Filtering selects the scales, while quantization selects the values.

\paragraph{Selecting spatial and temporal scales.}
Resampling without aliasing requires the removal of unresolved frequencies.
A spatial cutoff limits the retained wavenumbers.
A temporal cutoff limits how rapidly their coefficients change.
Both are required when space and time are resampled.

Sweeping provides a physical relation between the two cutoffs.
Appendix~\ref{app:turbulence-characteristics} describes how large motions
sweep smaller structures past fixed positions.
The temporal width of a shell then scales approximately as
$\urms k$~\citep{wilczek2012wavenumber}.
Let $k_0=2\pi/\domside$ be the fundamental spatial wavenumber.
Let $M$ set the number of samples per spatial direction and in time.
The operator $\mathcal P_M$ uses the coupled cutoffs
\begin{equation}
 k_c(M)=k_0\left(\frac M2-1\right),\qquad
 \omega_c(M)=\alpha\urms k_0\frac M2,\qquad
 \Delta t_M=\frac{\pi}{\omega_c(M)}.
 \label{eq:app-filter-cutoffs}
\end{equation}
Here $\omega$ is angular temporal frequency.
The dimensionless constant $\alpha$ sets the temporal bandwidth for a chosen
energy-retention level near the spatial cutoff.

\paragraph{Choosing the temporal interval.}
The physical domain is periodic in space, but a finite observation interval
has no equivalent temporal boundary condition.
A discrete Fourier transform repeats each transformed interval periodically.
If its endpoint frames differ, their artificial join adds high-frequency content.
A four-dimensional FFT therefore imposes a temporal join when it is applied
directly to a short movie.

The temporal transform uses the full record before evaluating a shorter
interior window.
This moves the join away from the selected times without removing all
boundary influence.
For identical intervals and cutoffs, a four-dimensional FFT and separate
spatial and temporal transforms are mathematically equivalent.
Their cutoffs remain distinct because $\alpha\urms$ converts wavenumber to
frequency in~(\ref{eq:app-filter-cutoffs}).

\paragraph{Filtering before resampling.}
For the full record, define
$\omega_\ell=2\pi\ell/(J\dtsnap)$, where $\ell$ indexes signed frequency bins.
The temporal Fourier coefficients and the filtered field are
\begin{align}
 \widetilde{\uvel}_{N,\sigma}(\wave,\omega_\ell)
 &=\frac1J\sum_{j=0}^{J-1}
   \uhat_{N,\sigma}(\wave,t_j)e^{-i\omega_\ell(t_j-t_0)},\notag\\
 (\mathcal P_M\uvel_{N,\sigma})(\pos,t)
 &=\sum_{\substack{\wave\in\mathcal K_N\\|\wave|\le k_c(M)}}
   \sum_{|\omega_\ell|<\omega_c(M)}
   \widetilde{\uvel}_{N,\sigma}(\wave,\omega_\ell)
   e^{i\wave\cdot\pos+i\omega_\ell(t-t_0)}.
 \label{eq:app-filtered-field}
\end{align}
The retained modes lie below the Nyquist limits $\pi M/\domside$ in space
and $\pi/\Delta t_M$ in time.
Applying the cutoffs before resampling prevents discarded modes from appearing
as false larger or slower motion~\citep{karban2022aliasing}.

Evaluating~(\ref{eq:app-filtered-field}) at $M^3$ positions and $M$ times
gives an $M\times M\times M\times M\times3$ array.
Its axes represent time, three spatial coordinates, and velocity components.
For fixed $\alpha$, $\urms$, and $\domside$, the window length
$T_w=M\Delta t_M=\domside/(\alpha\urms)$ is independent of $M$.
The half-open window spans $(M-1)\Delta t_M$ between its endpoint samples.

\paragraph{Quantizing the retained values.}
Quantization applies the rounding map to each retained velocity component:
\begin{equation}
 (\mathcal Q_q\uvel)_i(\pos,t)=r_q\bigl(u_i(\pos,t)\bigr),
 \qquad i=1,2,3.
 \label{eq:app-record-quantization}
\end{equation}
The format $q$ specifies saved values independently of solver precision $\sigma$.
During evolution, rounded values affect later updates.
During recording, they do not feed back into the trajectory.

When filtering precedes quantization, the represented record is
$\uvel^{\mathrm{rec}}_{N,\sigma,M,q}
=\mathcal Q_q\mathcal P_M\uvel_{N,\sigma}$.
The order must be specified because
$\mathcal Q_q\mathcal P_M\uvel_{N,\sigma}$ and
$\mathcal P_M\mathcal Q_q\uvel_{N,\sigma}$ generally differ.
Quantization can add variations outside the retained band.
Filtering afterward removes their out-of-band part.
Quantizing afterward retains the full rounding distortion in the saved values.

\clearpage
\section{Neural Representations}
\label{app:neural-framework}

Appendix~\ref{app:dns-framework} leaves us with a recorded velocity field
$\uvel^{\mathrm{rec}}$, sampled at discrete spatial locations and times.
A neural representation replaces this collection of samples with a
parameterized function. We first describe this approximation, then introduce
two ways of sharing parameters across time, and finally define the two
representation families used in our experiments.

\subsection{From velocity samples to a function}
\label{app:neural-approximation}

Let $\uvel^{\mathrm{rec}}(\pos_a,t_j)$ denote the recorded velocity at spatial
position $\pos_a$ and time $t_j$. The record contains $n_x$ spatial positions
and $n_t$ time points. For an $M^4$ record, $n_x=M^3$ and $n_t=M$.
A neural representation maps each space--time coordinate to the three velocity
components,
\begin{equation}
 f(\pos_a,t_j)
 =\bigl(f_1(\pos_a,t_j),f_2(\pos_a,t_j),f_3(\pos_a,t_j)\bigr)
 \approx \uvel^{\mathrm{rec}}(\pos_a,t_j).
 \label{eq:app-neural-values}
\end{equation}
Holding $t_j$ fixed and varying $\pos_a$ reconstructs a three-dimensional
snapshot of the flow. Holding $\pos_a$ fixed and varying $t_j$ instead gives
the velocity history at a single spatial location.

The function is determined by a set of learned scalar parameters. For a
multilayer network, these are the network weights and biases. For a Gaussian
field, they additionally include the locations, shapes, and velocity
coefficients of the basis functions. The architecture determines how these
parameters are used to map coordinates to velocity values.

Importantly, the full record is available during fitting. The task is therefore
one of \emph{reconstruction}, not prediction. Prediction would require
estimating states that were not observed during training. Here, evaluating a
fitted representation simply returns an approximation to a recorded state; it
does not advance an initial condition through the Navier--Stokes equations.

\paragraph{Fitting recorded velocities.}
Coordinate-based signal representations are commonly fitted by minimizing mean
squared error (MSE)~\citep{sitzmann2020implicit}. Giving equal weight to all
positions, times, and velocity components gives
\begin{equation}
 \mathcal L(f)=\frac{1}{3n_xn_t}
 \sum_{j=0}^{n_t-1}\sum_{a=1}^{n_x}
 \bigl\|f(\pos_a,t_j)-\uvel^{\mathrm{rec}}(\pos_a,t_j)\bigr\|_2^2.
 \label{eq:app-neural-mse}
\end{equation}
Training adjusts the representation parameters to reduce this average squared
velocity error.

Let $e=f-\uvel^{\mathrm{rec}}$ denote the reconstruction error. On a uniform
spatial grid, Parseval's identity gives
$\mathcal L(f)=(3n_t)^{-1}\sum_j\sum_{\wave}|\widehat e(\wave,t_j)|^2$
when the spatial Fourier transform is normalized by $1/n_x$. In this sense,
MSE measures the total energy of the reconstruction error across spatial
modes.

\subsection{Sharing parameters across time}
\label{app:neural-sharing}

The same samples and the same reconstruction loss can be paired with different
ways of distributing parameters across time. Let $B$ denote the total number
of stored scalar parameters available for the complete movie, and let
$|\theta|$ denote the size of a parameter set. At a fixed numerical precision,
$B$ also determines the storage budget.

\paragraph{Joint fitting.}
Joint fitting uses a single function $f_\theta(\pos,t)$ that takes both space
and time as input. The same parameters are shared across every frame, with
$|\theta|\leq B$. Changing $t$ changes the represented spatial field, but all
recorded times constrain the same parameter set.

\paragraph{Independent fitting.}
Independent fitting assigns a separate spatial function
$f_{\theta_j}(\pos)$ to each recorded time. The total parameter budget is
shared across these models,
$\sum_j|\theta_j|\leq B$. With an approximately uniform allocation, each frame
therefore receives about $B/n_t$ parameters. In this case, the frame index
selects an entirely separate parameter set rather than entering the function as
a coordinate.

These two procedures differ only in how information is shared across the
record. Independent fitting shares no parameters between frames. Joint fitting
treats the entire movie as a single function of space and time. 

\subsection{Two representation families}
\label{app:neural-families}

The way parameters are shared across time is separate from the choice of
function family. Continuous representations can be built from a variety of architectures, e.g., multilayer
decoders~\citep{park2019deepsdf}, sinusoidal networks
~\citep{sitzmann2020implicit}, Fourier-feature networks
~\citep{tancik2020fourier}, multiplicative-filter networks
~\citep{fathony2021multiplicative}, multiresolution feature grids
~\citep{muller2022instant}, or Gaussian basis-function networks
~\citep{poggio1990networks}. We focus on SIRENs and additive Gaussian fields
because they distribute their representational capacity in fundamentally
different ways. SIREN uses a single coordinate network across the full domain,
whereas the Gaussian field constructs the velocity field from localized,
adaptive primitives.

\paragraph{SIREN.}
A SIREN alternates affine transformations with sinusoidal activations
~\citep{sitzmann2020implicit}. Its hidden layers take the form
\begin{equation}
 h_{\ell+1}=\sin(W_\ell h_\ell+b_\ell),
 \label{eq:app-siren-layer}
\end{equation}
followed by a final affine layer that outputs the three velocity components.
For joint fitting, the network receives the four coordinates $(\pos,t)$.
Independent models receive only the three spatial coordinates $\pos$.

\paragraph{Additive Gaussian fields.}
Gaussian splatting represents a field using learned Gaussian
primitives~\citep{kerbl2023gaussian}. We adopt the same basic idea for velocity
fields and use the additive form
\begin{equation}
 f_j^{\mathrm G}(\xi)
 =\sum_{r=1}^{n_{\mathrm G}} \mathbf a_{rj}
 \exp\!\left[-\tfrac12
 \|L_{rj}(\xi-\boldsymbol\mu_{rj})\|_2^2\right]
 w_{rj}(\xi-\boldsymbol\mu_{rj}),
 \label{eq:app-gaussian-field}
\end{equation}
where $\boldsymbol\mu_{rj}$ specifies the primitive center, $L_{rj}$ its
anisotropic shape, $\mathbf a_{rj}\in\mathbb R^3$ its velocity coefficient,
and $w_{rj}$ a compact taper. The frame index is included here to allow these
attributes to vary when required by the fitting procedure. The primitive
contributions are summed directly, rather than combined through the opacity
compositing used in image rendering. The resulting model can be viewed as a
generalized Gaussian basis-function network with a three-component linear
output~\citep{poggio1990networks}.


\clearpage
\section{Experiments: Numerical Modeling and Representation}
\label{app:numerical-modeling-representation}
\label{app:numerical-setup}

Section~\ref{sec:numerical-experiments} considers four ways in which a velocity
record can change. Resolution $N$ and arithmetic precision $\sigma$ affect how
the trajectory is computed, whereas filtering $\mathcal P_M$ and quantization
$\mathcal Q_q$ act on the record after the simulation. Here we describe these
experiments using the solver and recording notation from
Appendix~\ref{app:dns-framework}, and report the measurements and storage
requirements for all records in Figure~\ref{fig:numerical-effects-on-turbulence}.

\subsection{Numerical modeling}

\paragraph{Physical problem and spatial discretization.}
We solve~(\ref{eq:app-ns}) on $[0,2\pi)^3$ with periodic boundaries and zero
mean velocity. Large-scale forcing sustains the turbulent motion. After every
full time step, we rescale the Fourier coefficients in shells
$\operatorname{round}(|\wave|)=1$ and $2$ so that their shell energies are
$0.8$ and $0.3$, respectively. The forcing is deterministic and introduces no
preferred direction.

On each $N^3$ grid, the nonlinear term is evaluated in rotational form,
$\mathsf P_{\wave}\widehat{\uvel\times\vort}$, which is equivalent to the
projected advective term in~(\ref{eq:app-spectral-evolution}). Products are
formed in physical space. We retain modes satisfying
$|k_i|\leq h_N=\lfloor N/3\rfloor$ in each direction and apply the
divergence-free projection to the nonlinear term. The retained region is
therefore a cube in Fourier space. We report $k_{\max}=h_N$, the radius of the
largest complete sphere contained in this cube. Corner modes extend beyond
this radius and are still included in the scalar statistics.

\paragraph{Time integration.}
Let $G(U)$ denote the dealiased, projected nonlinear term and let
$H=e^{-\visc|\wave|^2\dtsolv/2}$ act independently on each Fourier mode. The
third-order integrating-factor Runge--Kutta update is
\begin{align}
 U_a&=H^2U^n+\dtsolv H^2G(U^n),\notag\\
 U_b&=HU^n+\tfrac14\dtsolv HG(U^n)
              +\tfrac14\dtsolv H^{-1}G(U_a),\notag\\
 U_*&=H^2U^n+\tfrac16\dtsolv H^2G(U^n)
              +\tfrac16\dtsolv G(U_a)+\tfrac23\dtsolv HG(U_b).
 \label{eq:app-experimental-rk}
\end{align}
We then rescale the forced shells and remove the zero mode to obtain
$U^{n+1}$. Table~\ref{tab:numerical-inputs} summarizes the two physical
setups. Each comparison uses one initial realization for each setting.

\begin{table}[htbp]
\centering\normalsize
\caption{Physical and recording setup.}
\label{tab:numerical-inputs}
\setlength{\tabcolsep}{5pt}
\renewcommand{\arraystretch}{1.15}
\begin{tabular*}{\linewidth}{@{\extracolsep{\fill}}lll@{}}
\toprule
Quantity & Resolution comparison & Precision comparison \\
\midrule
Grid size $N$ & $32,64,128,256,512,1024$ & $128$ \\
Solver precision $\sigma$ & FP32 & FP64, FP32, FP16, FP8, FP4 \\
Viscosity $\visc$ & $0.00065$ & $0.010421385668748454$ \\
Nominal step $\dtsolv$ & $0.00040015018141313906$ & $0.004701911987471931$ \\
Steps between snapshots & $6$ & $2$ \\
Snapshot interval $\dtsnap$ & $0.0024009010884788345$ & $0.009403823974943862$ \\
Scheduled snapshots $J$ & $7{,}562$ & $2{,}362$ \\
First-to-last span $(J-1)\dtsnap$ & $18.15321312999$ & $22.20242840484$ \\
Update implementation & Native FP32 & Compensated, format-rounded \\
\bottomrule
\end{tabular*}
\end{table}

\paragraph{Resolution.}
The existing $1024^3$ simulation provides the reference trajectory. Its initial
random solenoidal field was generated with seed $20240617$, with a spectrum
proportional to $k^4e^{-2(k/2)^2}$ and initial $\urms=1$. The field was evolved
for about eight large-eddy turnover times before recording began. The first
archived state defines time zero for the resolution comparison.

For every smaller $N$, we take that same state and retain the modes satisfying
$|k_i|\leq h_N$. Because the stored forward FFT is unnormalized, the
coefficients are multiplied by $(N/1024)^3$ when transferred to the smaller
grid. This accounts for the FFT convention while preserving the physical mode
amplitudes. We do not renormalize the energy or discard an additional
transient. All simulations use the same native FP32 update, viscosity, forcing,
time step, and snapshot times. Only $N$ and the corresponding discrete
operators change.

The measured Reynolds number can therefore differ even though the physical
inputs are fixed. The coarse-grid results should be interpreted as outcomes of
under-resolved calculations, not as additional fully resolved DNS references.

\paragraph{Precision.}
The precision comparison begins from a single archived, equilibrated $128^3$
field at the higher viscosity listed in Table~\ref{tab:numerical-inputs}.
Every format receives the same initial field after rounding by $r_\sigma$.
For the formats that evolve successfully, we run $2{,}048$ nominal warm-up
steps, reset the recording clock to zero, and then save states at the intervals
listed above. The formats themselves are defined in
Table~\ref{tab:numerical-formats}.

All formats use an increment form of~(\ref{eq:app-experimental-rk}) with
compensated accumulation. A second array stores the accumulation residual
between updates. It uses the same format as the velocity state and is rescaled
together with that state during forcing. This implementation is separate from
the native FP32 solver used in the resolution study. The precision comparison
therefore varies $\sigma$ within one solver implementation, while the
resolution comparison varies $N$ within the other.

Each real addition, subtraction, multiplication, division, and square root is
followed by $r_\sigma$. Complex multiplication uses four rounded real
products and two rounded sums. Fourier transforms use radix-two butterflies
with rounding after every butterfly, and the forward transform divides by two
at each stage. Reductions use pairwise rounded sums. Wavevectors, time-step
coefficients, and forcing targets are rounded as well. Exponential factors and
Fourier phase constants are evaluated in binary64 and then rounded. The
compensated update uses \texttt{expm1} to form small viscous increments.
Binary64 is used to emulate the lower-precision formats.

\subsection{Representation}

\paragraph{Floating-point values and recorded velocities.}
The same map $r$ is used for arithmetic rounding and for encoding the recorded
values. All conversions round to the nearest representable value, with ties
to even, and retain subnormal values. FP64, FP32, and FP16 use the standard
binary IEEE formats~\citep{ieee2019floating}. For FP8 E4M3 and FP4 E2M1, we use
finite-value encodings~\citep{rouhani2023microscaling}
and saturate values outside their finite range. These choices are part of the
experiment. 

\begin{table}[htbp]
\centering\normalsize
\caption{Floating-point formats used for computation and recording.}
\label{tab:numerical-formats}
\setlength{\tabcolsep}{5pt}
\renewcommand{\arraystretch}{1.12}
\begin{tabular*}{\linewidth}{@{\extracolsep{\fill}}llrrl@{}}
\toprule
Label & Encoding & Bits per value & Largest finite magnitude & Overflow rule \\
\midrule
FP64 & binary64 & 64 & $1.798\times10^{308}$ & IEEE infinity \\
FP32 & binary32 & 32 & $3.403\times10^{38}$ & IEEE infinity \\
FP16 & binary16 & 16 & $65{,}504$ & IEEE infinity \\
FP8 & E4M3 & 8 & $448$ & Finite saturation \\
FP4 & E2M1 & 4 & $6$ & Finite saturation \\
\bottomrule
\end{tabular*}
\end{table}

At each saved time in the precision experiment, we reconstruct the physical
velocity from the evolved Fourier state using a common binary64 inverse
transform. Each velocity component is then encoded in the corresponding format
$\sigma$. The recorded precision experiment therefore includes both the effect
of arithmetic during the simulation and the final encoding of the output.

The quantization experiment isolates the second of these effects. We apply
$\mathcal Q_q$ once to the completed FP64 trajectory using each of the same
five formats. FP64 quantization is a bitwise no-op and therefore shares the
reference files. The low-precision encoders round directly from binary64,
avoiding an intermediate FP32 conversion near rounding boundaries. Every
encoded archive was decoded and checked during writing. 

\paragraph{Space--time filtering.}
The filtering experiment applies $\mathcal P_M$ from
(\ref{eq:app-filtered-field}) to $\uvel_{1024,\mathrm{FP32}}$ for
$M=32,64,128,256,512$. The spatial mask retains
$|\wave|\leq M/2-1$, while the temporal cutoff is
$|\omega|<\alpha\urms M/2$. We calibrate $\alpha$ once from the reference
trajectory and use the same value for every $M$.

The calibration is chosen to retain \textbf{99\% of the temporal spectral
energy} in a pooled, normalized shell-frequency distribution. This should not
be interpreted as retaining 99\% of the total DNS energy at every $M$.
Frequency widths are estimated from up to 512 sampled modes per shell, selected
with seed $20260751$. Welch spectra~\citep{welch1967power} use Hann windows of 1,024 samples with
50\% overlap for shells $1$--$63$, and windows of 512 samples for shells
$80,100,120,127$. For the higher shells, we use two 1,024-frame intervals
starting at frames 1,024 and 5,514. The width estimate removes the spectral
broadening introduced by the window.

Before estimating the widths, we linearly interpolate across
$|\omega|<2\Delta\omega$, where $\Delta\omega$ is the Welch frequency
spacing. The pooled distribution contains shells $16$--$63$ together with the
four higher shells, with each shell rescaled by its measured width. Its
two-sided 99\% quantile is $x_*=2.455$. Taking the largest interpolated
width-to-wavenumber ratio at $k=16,32,64,128$ gives
$c_{\mathrm{edge}}=1.03358067410$, and therefore
\begin{equation}
 \alpha=\frac{x_*c_{\mathrm{edge}}}{\urms}=2.31005299598,
 \qquad \urms=1.09843391443,
 \qquad \alpha\urms=2.53744055491.
 \label{eq:app-filter-calibration}
\end{equation}

For each retained spatial mode, we transform the complete 7,562-frame time
series, keep only the selected frequency bins, and evaluate the resulting
Fourier series at
$t_j=9.07780701554+(j-M/2)\Delta t_M$, where
$\Delta t_M=\pi/(\alpha\urms M/2)$. All five filtered records cover the same
half-open interval
$[7.83971194075,10.31590209032)$, with duration
$T_w=2.47619014957$.

Temporal transforms and evaluation use binary64 arithmetic. For
$M\leq256$, the construction stores intermediate coefficients in complex64
and outputs three FP32 velocity components at each of the $M^4$ coordinates.
For $M=512$, the same projection is evaluated using streamed matrix products.
Direct Fourier-series checks confirm the numerical evaluation. 

\subsection{Measurements and numerical outcomes}
\label{app:numerical-outcomes}

\paragraph{Common observation window.}
The unfiltered records are measured over the same nominal time interval as the
filtered movies. This interval contains 1,031 native snapshots for the
resolution comparison and 263 snapshots for the precision and quantization
comparisons.

For these records, spatial statistics are computed from 16 evenly spaced
snapshots. The $1024^3$ reference uses 81 snapshots for its energy spectrum,
16 for the gradient moments, and all 1,031 recorded dissipation values for
enstrophy. For the filtered records, all $M$ frames are used. At
$M=256,512$, cached gradient moments were checked against the common
derivative operator before being reused. All reported means therefore refer to
this finite observation window and to a single realization.

\paragraph{Spatial and temporal estimators.}
Spatial derivatives are evaluated spectrally in binary64 using the actual
stored fields. Nyquist derivative multipliers are set to zero, as required for
the real trigonometric interpolant. We apply neither smoothing nor a
divergence-free projection before computing the diagnostics. Enstrophy is
obtained from the curl, while $S$ and $F$ are normalized separately within
each direction and frame before averaging.

For the reference simulation, enstrophy is computed as
$\overline\diss/(2\visc)$. A direct Fourier-space check on the 16 frames used
for the gradient moments confirms negligible divergence. Spectra are shown
only over their complete-shell range, whereas the scalar statistics include
all stored modes.

Correlation times use all available samples in the observation window and up
to 512 Fourier modes per shell, chosen using seed $20260910$. We average the
real mode products over all valid time pairs, without subtracting the mean or
applying a taper, and normalize by the zero-lag value. The first crossing of
$e^{-1}$ is found by linear interpolation, considering lags up to half the
record length. Shells without a measured crossing are omitted from the figure.
Table~\ref{tab:numerical-all-records} reports two representative shells, while
the accompanying data include every measured $E(k)$ and $\tau_e(k)$.

\paragraph{Incompressibility.}
To compare fields with different gradient strengths, we also report
\begin{equation}
 d_\nabla=
 \left[\frac{\overline{\langle(\nabla\cdot\uvel)^2\rangle}}
 {2\overline\enst+
  \overline{\langle(\nabla\cdot\uvel)^2\rangle}}\right]^{1/2}.
 \label{eq:app-numerical-divergence}
\end{equation}
By the periodic identity in
Appendix~\ref{app:turbulence-characteristics}, the denominator is
$\overline{\langle|\nabla\uvel|^2\rangle}$. Values close to zero mean that
divergence is small relative to the overall velocity gradients. This is a
check of one physical constraint; it does not by itself show that the flow is
resolved or statistically stationary.

\begin{table}[!t]
\centering\normalsize
\caption{Measured records and stored payloads underlying Figure~\ref{fig:numerical-effects-on-turbulence}.}
\label{tab:numerical-all-records}
\setlength{\tabcolsep}{3.5pt}
\renewcommand{\arraystretch}{1.12}
\begin{tabular*}{\linewidth}{@{\extracolsep{\fill}}lrrrrrrrr@{}}
\toprule
Record & $\overline K$ & $\overline\enst$ & $\overline S$ & $\overline F$ & $d_\nabla$ & $\tau_e(8)$ & $\tau_e(24)$ & TB \\
\midrule
\multicolumn{9}{@{}l}{\textbf{Resolution}} \\
$N=1024$ & 1.795 & 323.20 & -0.562 & 7.005 & $2.91\!\times\!10^{-6}$ & 0.206 & 0.058 & 28.954457 \\
$N=512$ & 1.734 & 291.14 & -0.532 & 7.055 & $2.56\!\times\!10^{-6}$ & 0.213 & 0.059 & 3.608717 \\
$N=256$ & 1.824 & 330.24 & -0.354 & 5.160 & $2.53\!\times\!10^{-6}$ & 0.203 & 0.058 & 0.456393 \\
$N=128$ & 1.771 & 314.76 & -0.143 & 3.676 & $2.61\!\times\!10^{-6}$ & 0.235 & 0.056 & 0.056384 \\
$N=64$ & 2.235 & 448.19 & -0.037 & 3.126 & $5.99\!\times\!10^{-6}$ & 0.171 & --- & 0.007383 \\
$N=32$ & 8.695 & 828.27 & -0.008 & 3.007 & $7.39\!\times\!10^{-4}$ & 0.112 & --- & 0.000880 \\
\addlinespace[3pt]
\multicolumn{9}{@{}l}{\textbf{Filtering}} \\
$M=512$ & 1.795 & 320.99 & -0.548 & 6.666 & $2.94\!\times\!10^{-6}$ & 0.206 & 0.058 & 0.824634 \\
$M=256$ & 1.794 & 290.12 & -0.474 & 5.462 & $3.09\!\times\!10^{-6}$ & 0.207 & 0.058 & 0.051540 \\
$M=128$ & 1.780 & 191.02 & -0.381 & 4.344 & $3.78\!\times\!10^{-6}$ & 0.207 & 0.058 & 0.003221 \\
$M=64$ & 1.722 & 81.89 & -0.329 & 3.807 & $5.70\!\times\!10^{-6}$ & 0.207 & 0.060 & 0.000201 \\
$M=32$ & 1.609 & 27.60 & -0.312 & 3.555 & $9.61\!\times\!10^{-6}$ & 0.209 & --- & 0.000013 \\
\addlinespace[3pt]
\multicolumn{9}{@{}l}{\textbf{Precision}} \\
FP64 & 1.617 & 19.25 & -0.420 & 4.374 & $2.91\!\times\!10^{-15}$ & 0.245 & 0.087 & 0.118883 \\
FP32 & 1.612 & 19.17 & -0.429 & 4.355 & $2.94\!\times\!10^{-7}$ & 0.244 & 0.087 & 0.059442 \\
FP16 & 1.572 & 17.84 & -0.456 & 4.313 & $3.26\!\times\!10^{-3}$ & 0.246 & 0.090 & 0.029721 \\
FP8 & 1.601 & 21.47 & -0.176 & 3.724 & $2.76\!\times\!10^{-1}$ & $\infty$ & $\infty$ & 0.014860 \\
FP4 & \multicolumn{7}{c}{No evolved record} & --- \\
\addlinespace[3pt]
\multicolumn{9}{@{}l}{\textbf{Quantization}} \\
FP64 & 1.617 & 19.25 & -0.420 & 4.374 & $2.91\!\times\!10^{-15}$ & 0.245 & 0.087 & 0.118883 \\
FP32 & 1.617 & 19.25 & -0.420 & 4.374 & $2.69\!\times\!10^{-7}$ & 0.245 & 0.087 & 0.059442 \\
FP16 & 1.617 & 19.25 & -0.420 & 4.374 & $2.21\!\times\!10^{-3}$ & 0.245 & 0.087 & 0.029721 \\
FP8 & 1.616 & 22.14 & -0.258 & 4.090 & $2.43\!\times\!10^{-1}$ & 0.245 & 0.085 & 0.014860 \\
FP4 & 1.631 & 74.82 & -0.030 & 3.900 & $4.77\!\times\!10^{-1}$ & 0.243 & 0.069 & 0.007430 \\
\addlinespace[3pt]
\bottomrule
\end{tabular*}
\end{table}

All quantities use the dimensionless units of the simulation. The storage
column refers to the full archived record, not only to the common observation
window. A dash indicates that the corresponding shell or record is
unavailable.

The FP8 precision field is a verified fixed point, so its nonzero shells have
$\tau_e=\infty$, although its spatial statistics are still well defined. The
FP4 precision simulation cannot initialize because both the nominal time step
and the viscosity round to zero, and therefore produces no evolving record.
Quantization is different: all five formats yield finite recorded fields.

\newpage
\paragraph{Flow scales and resolution checks.}
Table~\ref{tab:numerical-flow-scales} reports the usual flow and sampling
scales for the evolving records. We compute
$L=(3\pi/4\overline K)\sum_{k\geq1}\overline E(k)/k$ and use the definitions
in~(\ref{eq:app-turbulent-scales}). When divergence is nonzero, we define the
strain-based dissipation consistently as
$\overline\diss=2\visc[\overline\enst+
\overline{\langle(\nabla\cdot\uvel)^2\rangle}]$.

The table uses window-averaged quantities. For strongly distorted records,
these values should be regarded as diagnostic scale estimates rather than as
evidence of developed turbulence. In particular, $k_{\max}\eta<1$ for
$N\leq512$, so using the same forcing and viscosity does not make the coarse
simulations spatially resolved.

\begin{table}[!htbp]
\centering\normalsize
\caption{Flow scales estimated on the observation window.}
\label{tab:numerical-flow-scales}
\setlength{\tabcolsep}{4pt}
\renewcommand{\arraystretch}{1.12}
\begin{tabular*}{\linewidth}{@{\extracolsep{\fill}}lrrrrrrr@{}}
\toprule
Record & $\urms$ & $\overline\diss$ & $\intscale$ & $\Relam$ & $k_{\max}\eta$ & $\dtsnap/\tau_\eta$ & $(J-1)\dtsnap/T_L$ \\
\midrule
$N=1024$ & 1.094 & 0.4202 & 1.393 & 280.5 & 1.724 & 0.061 & 14.26 \\
$N=512$ & 1.075 & 0.3785 & 1.426 & 285.5 & 0.882 & 0.058 & 13.68 \\
$N=256$ & 1.103 & 0.4293 & 1.379 & 281.9 & 0.427 & 0.062 & 14.51 \\
$N=128$ & 1.087 & 0.4092 & 1.411 & 280.5 & 0.214 & 0.060 & 13.98 \\
$N=64$ & 1.221 & 0.5826 & 1.096 & 296.5 & 0.098 & 0.072 & 20.22 \\
$N=32$ & 2.408 & 1.0767 & 0.487 & 848.6 & 0.040 & 0.098 & 89.77 \\
$N=128$, FP64 & 1.038 & 0.4011 & 1.562 & 64.6 & 1.721 & 0.058 & 14.76 \\
$N=128$, FP32 & 1.037 & 0.3995 & 1.565 & 64.5 & 1.723 & 0.058 & 14.71 \\
$N=128$, FP16 & 1.024 & 0.3719 & 1.590 & 65.2 & 1.754 & 0.056 & 14.29 \\
\bottomrule
\end{tabular*}
\end{table}

All resolution simulations remain finite, although the $N=32$ run grows
strongly. Over the full monitored trajectory, its maximum
divergence-to-gradient ratio is $0.02775$, exceeding the campaign threshold of
$10^{-3}$. The smaller value reported in
Table~\ref{tab:numerical-all-records} refers only to the central observation
window. All runs with $N\geq64$ pass the divergence check.

Differences between grids over this finite window do not establish monotone
convergence. FP16 evolution remains finite and time dependent, but develops
measurable divergence.
For FP8, three update checks leave both the state and the compensation array
bitwise unchanged. The full frozen archive was therefore generated by
repeating this verified state at the scheduled times.

These results apply to the particular unscaled arithmetic algorithm used here.
They do not imply that a different scaling strategy or mixed-precision solver
would behave in the same way.

\subsection{Storage and reproducibility}

\paragraph{What is stored.}
The resolution archives contain only the retained real-FFT coefficients.
Each snapshot is stored as a complex64 array with shape
$3\times(2h_N+1)\times(2h_N+1)\times(h_N+1)$. Since each complex value
contains two four-byte real components, the payload of one complete record is
$24J(2h_N+1)^2(h_N+1)$ bytes. The $1024^3$ archive alone occupies
$28.954457$ decimal TB. This representation avoids storing zeros and redundant
Hermitian coefficients without introducing any further quantization of the
retained modes.

The precision and quantization archives instead store all three physical
velocity components at every grid point. For a format with $b$ bits per value,
the payload is $3JN^3b/8$ bytes. FP4 stores two values per byte. These are the
actual written payloads for FP16 and FP8 as well, even though the lower formats
are emulated using binary64 working arrays. Solver working memory and archived
record size are therefore separate quantities.

\begin{table}[!htbp]
\centering\normalsize
\caption{Distinct stored payloads of the records depicted in Figure~\ref{fig:numerical-effects-on-turbulence}.}
\label{tab:numerical-storage-total}
\begin{tabular*}{\linewidth}{@{\extracolsep{\fill}}lr@{}}
\toprule
Contribution & TB \\
\midrule
Resolution records, including the finite $N=32$ failure & 33.084213 \\
Precision records, including the frozen FP8 control & 0.222906 \\
Five filtered movies & 0.879608 \\
Additional quantized records, excluding the shared FP64 reference & 0.111453 \\
\midrule
\textbf{All distinct records} & $\mathbf{34.298181}$ \\
Computed records excluding the frozen FP8 and failed $N=32$ controls & 33.291379 \\
\bottomrule
\end{tabular*}
\end{table}

The tables use $1\,\mathrm{TB}=10^{12}$ bytes. Records depicted in Figure~\ref{fig:numerical-effects-on-turbulence} enter its accounting.
Their computed records contain $33.291$~TB after excluding the frozen FP8
control and the failed $N=32$ record. This includes under-resolved
calculations and is not a total of fully resolved DNS alone.
Including the depicted failure controls and the distinct filtered and
quantized records gives $34.298$~TB.

\paragraph{Reproduction records.}
The source manifests record array shapes, coordinate conventions, time steps,
viscosities, frame counts, and storage formats. They also identify the initial
states and solver versions using SHA-256 checksums. The reference simulation
used JAX 0.9.1, while the coarse-grid simulations used JAX 0.11.0.

Replaying one snapshot interval of the reference with the pinned solver gives
a relative Fourier-coefficient error of $8.2\times10^{-9}$. This confirms
short-time agreement, but not bitwise reproducibility across different
backends. The filtering manifests additionally store the cutoffs, sample
times, retained frequency bins, and construction checks.

\begin{figure}[p]
\centering
\includegraphics[width=\linewidth]{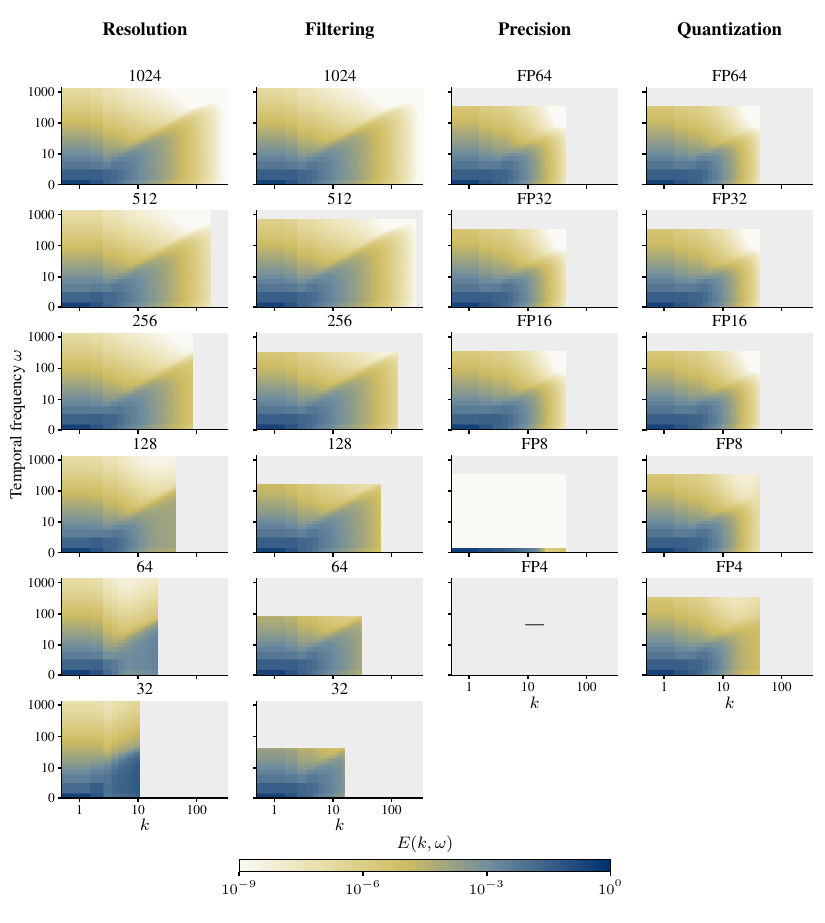}
\caption{\textbf{Joint spatial--temporal spectra of the numerical records.}
This adds the frequency distribution to the comparisons in
Figure~\ref{fig:numerical-effects-on-turbulence}. \emph{Resolution} changes the distribution even
at shared scales. \emph{Filtering} restricts its spatial and temporal extent.
For \emph{precision}, the frozen FP8 record retains zero-frequency energy,
while FP4 supplies no valid temporal spectrum.
Coarse \emph{quantization} broadens the small-scale temporal content.
All panels use the same absolute color scale. Gray denotes unrepresented
spatial or temporal frequencies. Spectra use up to 512 sampled Fourier modes
per shell, scaled to the measured full-shell energy. The temporal mean is
retained. The frequency axis is linear near zero and logarithmic above it.}
\label{fig:numerical-joint-spectrum}
\end{figure}

\clearpage
\section{Experiments: Neural Representations}
\label{app:neural-experiments}

Appendix~\ref{app:numerical-modeling-representation} establishes how the
computed and recorded flows differ. We now specify which part of each record
is supplied to a neural representation and how that representation is fitted.
We use the functions, loss, and parameter budget $B$ defined in
Appendix~\ref{app:neural-framework}.

\subsection{Learning targets}
\label{app:neural-targets}

\paragraph{A common observation window.}
We apply $\mathcal P_{64}$ to the resolution, precision, and quantization
records before fitting. This gives each comparison the same spatial cutoff,
temporal cutoff, and sample times. The filtering comparison instead varies
$M$, as listed in Table~\ref{tab:neural-targets}.
All targets retain three spatial coordinates and time.

\begin{table}[!ht]
\centering\normalsize
\caption{Records supplied to neural fitting.}
\label{tab:neural-targets}
\setlength{\tabcolsep}{5pt}
\renewcommand{\arraystretch}{1.15}
\begin{tabular*}{\linewidth}{@{\extracolsep{\fill}}lll@{}}
\toprule
Comparison & Supplied record $\uvel^{\mathrm{rec}}$ & Settings \\
\midrule
Resolution & $\mathcal P_{64}\uvel_{N,\mathrm{FP32}}$ & $N=128,256,512,1024$ \\
Filtering & $\mathcal P_M\uvel_{1024,\mathrm{FP32}}$ & $M=32,64,128$ \\
Precision & $\mathcal P_{64}\uvel_{128,\sigma}$ & $\sigma=\mathrm{FP64},\mathrm{FP32},\mathrm{FP16}$ \\
Quantization & $\mathcal P_{64}\mathcal Q_q\uvel_{128,\mathrm{FP64}}$ & $q=\mathrm{FP64},\mathrm{FP32},\mathrm{FP16},\mathrm{FP8},\mathrm{FP4}$ \\
\bottomrule
\end{tabular*}
\end{table}

The projection uses the calibrated cutoff and window in
Appendix~\ref{app:numerical-modeling-representation}.
At $M=64$, $k_c=31$, $\omega_c=81.1980977570$, and
$\Delta t_{64}=0.03869047109$.
The $64^4$ samples cover the same interval of length $T_w=2.47619014957$.
For $M=32$ and $128$, $\Delta t_M=T_w/M$ and $k_c=M/2-1$.
Each projection uses the complete source time series before evaluating this
window. The resulting arrays have shape $(M,M,M,M,3)$, ordered as
$(t,x,y,z,\text{velocity component})$, and store FP32 values.
This storage precision can mask differences between the finest quantization levels.

Quantization precedes projection: the learner receives a filtered version of
an already rounded record. No solenoidal projection is added.
The FP64 quantization target is the precision reference, and the
$N=1024$ resolution target is the $M=64$ filtering target.
These shared records are fitted once, giving 13 distinct targets.
The $N=32,64$ records and FP8/FP4 evolution outcomes are excluded from this
learning comparison. The numerical study retains them as under-resolution
or arithmetic-failure cases.

\begin{figure}[p]
\centering
\begin{minipage}{\linewidth}
\centerline{\includegraphics[width=.93\linewidth]{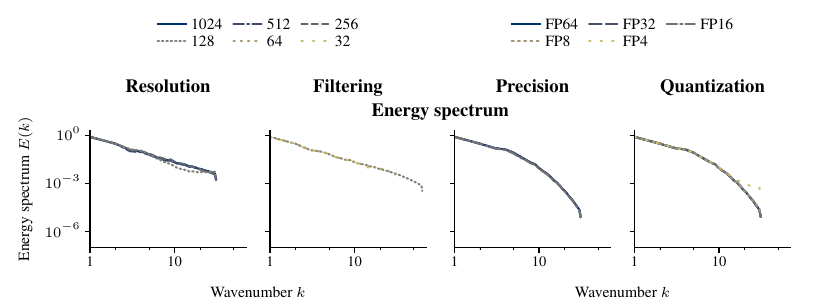}}
\par\smallskip
{\normalsize\noindent\captiona\ The \textbf{energy spectrum} retains differences across
source \emph{resolutions}, with less small-scale energy at $N=128$.
Wider \emph{filtering} bands recover more of the spectrum.
The evolving \emph{precision} records remain close, while FP4
\emph{quantization} still leaves excess small-scale energy after projection.\par}
\end{minipage}
\par\medskip
\begin{minipage}{\linewidth}
\centerline{\includegraphics[width=.93\linewidth]{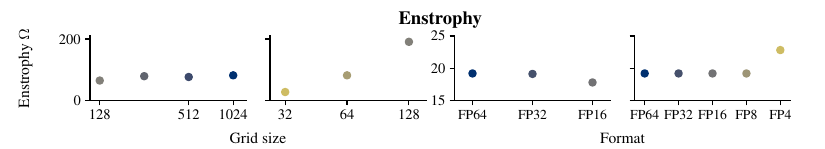}}
\par\smallskip
{\normalsize\noindent\captionb\ \textbf{Enstrophy} is lowest at $N=128$ in the
\emph{resolution} comparison and grows with wider \emph{filtering} bandwidth.
The FP16 \emph{precision} deficit remains.
FP4 \emph{quantization} still raises enstrophy, although projection removes
some of the excess present in the unfiltered record.\par}
\end{minipage}
\par\medskip
\begin{minipage}{\linewidth}
\centerline{\includegraphics[width=.93\linewidth]{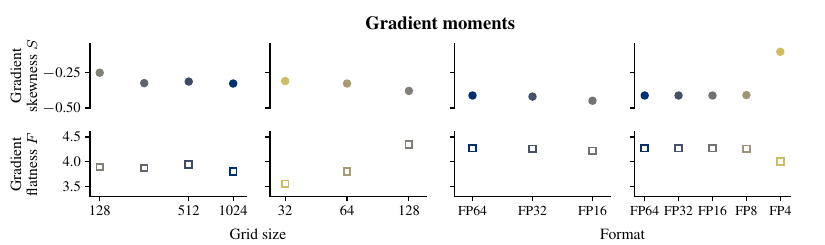}}
\par\smallskip
{\normalsize\noindent\captionc\ The \textbf{gradient moments} become more similar
across source \emph{resolutions} after projection.
Wider \emph{filtering} bands retain more negative skewness and higher flatness.
FP16 \emph{precision} retains more negative skewness than FP64, whereas
FP4 \emph{quantization} leaves weaker skewness.\par}
\end{minipage}
\par\medskip
\begin{minipage}{\linewidth}
\centerline{\includegraphics[width=.93\linewidth]{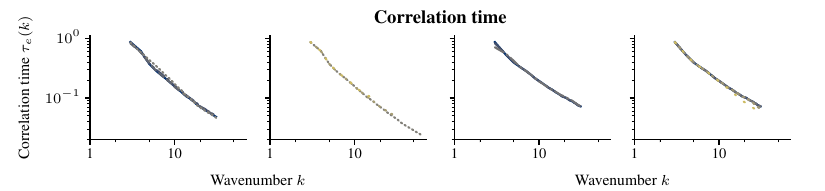}}
\par\smallskip
{\normalsize\noindent\captiond\ \textbf{Correlation times} remain scale dependent.
The $N=128$ \emph{resolution} record retains longer correlations at intermediate
scales. Different \emph{filtering} bandwidths largely agree on shared scales.
The evolving \emph{precision} records remain close, while FP4
\emph{quantization} still shortens small-scale correlations.\par}
\end{minipage}
\caption{Recorded motion supplied to neural fitting.}
\label{fig:training-data-characteristics}
\end{figure}

\paragraph{Effects of filtering.}
Figure~\ref{fig:training-data-characteristics} shows the motion supplied to fitting.
The \emph{energy spectra} and \emph{enstrophy} retain differences between
records, including the FP16 evolution deficit and the FP4 quantization excess.
The \emph{gradient moments} show a stronger change: the resolution records
become more similar after their finest scales are removed, while FP4
quantization still leaves much weaker skewness.
The \emph{correlation times} retain the shortening associated with FP4
quantization and the broad scale dependence of the source records.
Filtering therefore preserves the informative contrasts within the
retained band. The fitting experiments concern this motion.

\begin{figure}[p]
\centering
\begin{minipage}{\linewidth}
\centerline{\includegraphics[width=.93\linewidth]{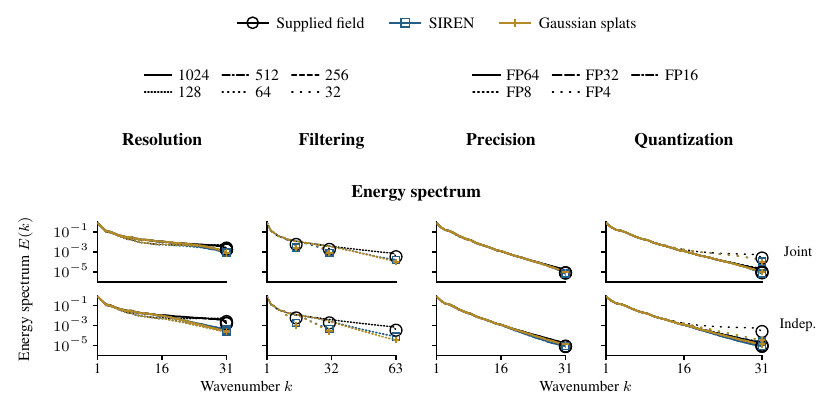}}
\par\smallskip
{\normalsize\noindent\captiona\ The \textbf{energy spectrum} retains the energy-containing
scales under both \emph{joint} and \emph{independent fitting}. Small-scale energy
is generally underestimated, especially by independent fits. Across
\emph{resolution} and \emph{filtering}, this removes supplied structure;
for coarse \emph{quantization}, it attenuates numerical excess.\par}
\end{minipage}
\par\medskip
\begin{minipage}{\linewidth}
\centerline{\includegraphics[width=.93\linewidth]{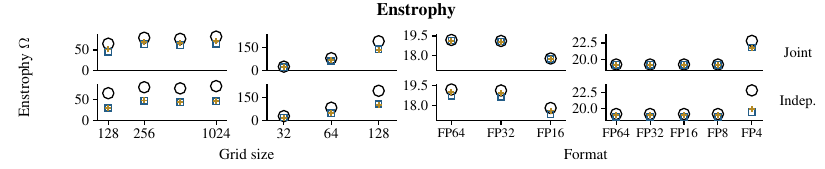}}
\par\smallskip
{\normalsize\noindent\captionb\ \textbf{Enstrophy} is generally lower under
\emph{independent fitting} than under \emph{joint fitting}. Deficits occur across
source \emph{resolutions} and grow with wider \emph{filtering} bandwidths.
On the FP4-quantized record, this reduction instead brings enstrophy closer
to the unquantized value.\par}
\end{minipage}
\caption{Joint and independent neural representations of the recorded flow.}
\label{fig:neural-representation-comparison}
\end{figure}

\begin{figure}[p]
\centering
\begin{minipage}{\linewidth}
\centerline{\includegraphics[width=.93\linewidth]{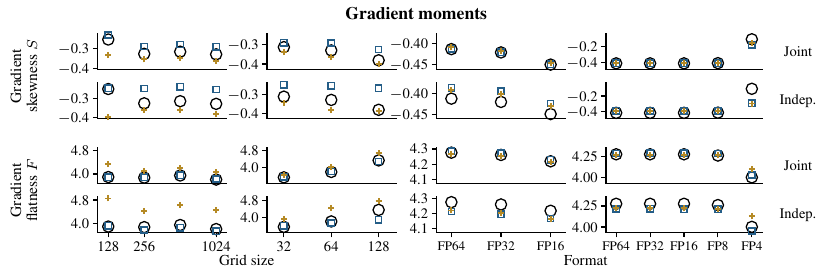}}
\par\smallskip
{\normalsize\noindent\captionc\ The \textbf{gradient moments} distinguish the network
families: across \emph{resolution} and \emph{filtering}, \emph{Gaussian splats}
have more negative skewness and higher flatness than \emph{SIRENs}.
\emph{Joint fitting} generally preserves the supplied gradient statistics
more closely than \emph{independent fitting}.\par}
\end{minipage}
\par\medskip
\begin{minipage}{\linewidth}
\centerline{\includegraphics[width=.93\linewidth]{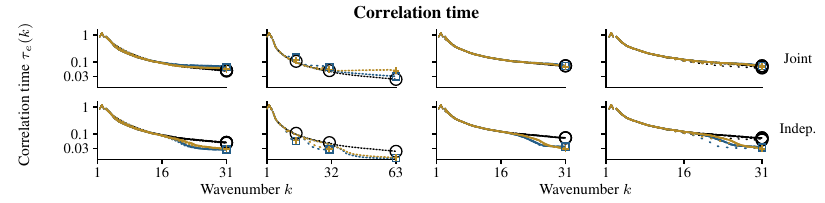}}
\par\smallskip
{\normalsize\noindent\captiond\ \textbf{Correlation times} expose a temporal trade-off.
\emph{Independent fits} generally decorrelate too rapidly at small scales,
whereas \emph{joint fits} can remain correlated too long. This tendency spans
\emph{resolution}, \emph{filtering}, \emph{precision}, and \emph{quantization};
the more accurate procedure depends on the record and network family.\par}
\end{minipage}
\par\medskip
{\normalsize\noindent Figure~\ref{fig:neural-representation-comparison} (continued).\par}
\end{figure}

\clearpage
\subsection{Absolute diagnostics of records and fits}
\label{app:neural-absolute-diagnostics}

Figure~\ref{fig:neural-representation-comparison} gives the complete comparison,
including the gradient moments omitted from the main figure.
Tables~\ref{tab:neural-diagnostics-resolution}--\ref{tab:neural-diagnostics-quantization}
list the absolute diagnostics for each supplied record and its four fits.
The supplied record is the reference for fitting, including when it contains
quantization distortion. The FP64 quantization block provides the corresponding
unquantized reference.
We use the same definitions and averaging as
Table~\ref{tab:numerical-all-records}, evaluated on the learning targets and
reconstructed movies. The incompressibility measure $d_\nabla$ is defined
in~(\ref{eq:app-numerical-divergence}). No divergence-free projection is applied
to the fitted fields. A dash marks a shell outside the retained band.
Shared targets are repeated under the relevant comparisons, without counting
them as additional fits. All values describe the runs plotted in the figure.

\begin{table}[!ht]
\centering\normalsize
\caption{Absolute diagnostics of supplied records and neural fits: resolution.}
\label{tab:neural-diagnostics-resolution}
\setlength{\tabcolsep}{3pt}
\renewcommand{\arraystretch}{1.06}
\begin{tabular*}{\linewidth}{@{\extracolsep{\fill}}llrrrrrrr@{}}
\toprule
Record & Representation & $\overline K$ & $\overline\enst$ & $\overline S$ & $\overline F$ & $d_\nabla$ & $\tau_e(8)$ & $\tau_e(24)$ \\
\midrule
$N=1024$ & Supplied & 1.722 & 81.89 & -0.329 & 3.807 & $5.70\!\times\!10^{-6}$ & 0.207 & 0.060 \\
 & SIREN, joint & 1.697 & 64.42 & -0.290 & 3.826 & $2.02\!\times\!10^{-1}$ & 0.207 & 0.074 \\
 & Gaussian, joint & 1.708 & 70.75 & -0.364 & 4.053 & $1.21\!\times\!10^{-1}$ & 0.207 & 0.063 \\
 & SIREN, indep. & 1.660 & 45.96 & -0.256 & 3.735 & $2.57\!\times\!10^{-1}$ & 0.207 & 0.031 \\
 & Gaussian, indep. & 1.662 & 46.03 & -0.383 & 4.450 & $1.76\!\times\!10^{-1}$ & 0.207 & 0.038 \\
\addlinespace[4pt]
$N=512$ & Supplied & 1.668 & 76.54 & -0.316 & 3.947 & $4.91\!\times\!10^{-6}$ & 0.213 & 0.062 \\
 & SIREN, joint & 1.648 & 61.41 & -0.279 & 3.958 & $1.96\!\times\!10^{-1}$ & 0.213 & 0.074 \\
 & Gaussian, joint & 1.656 & 66.67 & -0.347 & 4.195 & $1.19\!\times\!10^{-1}$ & 0.213 & 0.064 \\
 & SIREN, indep. & 1.612 & 43.42 & -0.242 & 3.855 & $2.57\!\times\!10^{-1}$ & 0.213 & 0.031 \\
 & Gaussian, indep. & 1.614 & 43.72 & -0.361 & 4.630 & $1.76\!\times\!10^{-1}$ & 0.213 & 0.038 \\
\addlinespace[4pt]
$N=256$ & Supplied & 1.752 & 79.19 & -0.327 & 3.874 & $5.06\!\times\!10^{-6}$ & 0.203 & 0.061 \\
 & SIREN, joint & 1.730 & 63.69 & -0.289 & 3.892 & $1.94\!\times\!10^{-1}$ & 0.203 & 0.073 \\
 & Gaussian, joint & 1.739 & 69.48 & -0.355 & 4.092 & $1.15\!\times\!10^{-1}$ & 0.203 & 0.064 \\
 & SIREN, indep. & 1.695 & 46.91 & -0.252 & 3.778 & $2.47\!\times\!10^{-1}$ & 0.203 & 0.030 \\
 & Gaussian, indep. & 1.699 & 47.66 & -0.360 & 4.430 & $1.69\!\times\!10^{-1}$ & 0.203 & 0.038 \\
\addlinespace[4pt]
$N=128$ & Supplied & 1.643 & 65.00 & -0.254 & 3.894 & $5.54\!\times\!10^{-6}$ & 0.235 & 0.059 \\
 & SIREN, joint & 1.618 & 45.95 & -0.230 & 3.890 & $2.34\!\times\!10^{-1}$ & 0.235 & 0.074 \\
 & Gaussian, joint & 1.626 & 51.26 & -0.334 & 4.337 & $1.49\!\times\!10^{-1}$ & 0.235 & 0.063 \\
 & SIREN, indep. & 1.588 & 29.73 & -0.249 & 3.923 & $2.52\!\times\!10^{-1}$ & 0.235 & 0.030 \\
 & Gaussian, indep. & 1.589 & 29.99 & -0.397 & 4.855 & $1.84\!\times\!10^{-1}$ & 0.235 & 0.036 \\
\bottomrule
\end{tabular*}
\end{table}

\begin{table}[!ht]
\centering\normalsize
\caption{Absolute diagnostics of supplied records and neural fits: filtering.}
\label{tab:neural-diagnostics-filtering}
\setlength{\tabcolsep}{3pt}
\renewcommand{\arraystretch}{1.06}
\begin{tabular*}{\linewidth}{@{\extracolsep{\fill}}llrrrrrrr@{}}
\toprule
Record & Representation & $\overline K$ & $\overline\enst$ & $\overline S$ & $\overline F$ & $d_\nabla$ & $\tau_e(8)$ & $\tau_e(24)$ \\
\midrule
$M=128$ & Supplied & 1.780 & 191.02 & -0.381 & 4.344 & $3.78\!\times\!10^{-6}$ & 0.207 & 0.058 \\
 & SIREN, joint & 1.760 & 139.82 & -0.325 & 4.293 & $2.13\!\times\!10^{-1}$ & 0.207 & 0.058 \\
 & Gaussian, joint & 1.757 & 133.88 & -0.399 & 4.675 & $1.70\!\times\!10^{-1}$ & 0.207 & 0.058 \\
 & SIREN, indep. & 1.738 & 105.78 & -0.269 & 3.880 & $2.58\!\times\!10^{-1}$ & 0.207 & 0.057 \\
 & Gaussian, indep. & 1.735 & 99.52 & -0.385 & 4.757 & $2.14\!\times\!10^{-1}$ & 0.207 & 0.057 \\
\addlinespace[4pt]
$M=64$ & Supplied & 1.722 & 81.89 & -0.329 & 3.807 & $5.70\!\times\!10^{-6}$ & 0.207 & 0.060 \\
 & SIREN, joint & 1.697 & 64.42 & -0.290 & 3.826 & $2.02\!\times\!10^{-1}$ & 0.207 & 0.074 \\
 & Gaussian, joint & 1.708 & 70.75 & -0.364 & 4.053 & $1.21\!\times\!10^{-1}$ & 0.207 & 0.063 \\
 & SIREN, indep. & 1.660 & 45.96 & -0.256 & 3.735 & $2.57\!\times\!10^{-1}$ & 0.207 & 0.031 \\
 & Gaussian, indep. & 1.662 & 46.03 & -0.383 & 4.450 & $1.76\!\times\!10^{-1}$ & 0.207 & 0.038 \\
\addlinespace[4pt]
$M=32$ & Supplied & 1.609 & 27.60 & -0.312 & 3.555 & $9.61\!\times\!10^{-6}$ & 0.209 & --- \\
 & SIREN, joint & 1.588 & 23.92 & -0.290 & 3.574 & $1.63\!\times\!10^{-1}$ & 0.211 & --- \\
 & Gaussian, joint & 1.594 & 24.52 & -0.338 & 3.682 & $1.08\!\times\!10^{-1}$ & 0.208 & --- \\
 & SIREN, indep. & 1.539 & 17.77 & -0.250 & 3.602 & $2.48\!\times\!10^{-1}$ & 0.204 & --- \\
 & Gaussian, indep. & 1.542 & 17.62 & -0.343 & 3.924 & $1.60\!\times\!10^{-1}$ & 0.203 & --- \\
\bottomrule
\end{tabular*}
\end{table}

\begin{table}[!ht]
\centering\normalsize
\caption{Absolute diagnostics of supplied records and neural fits: precision.}
\label{tab:neural-diagnostics-precision}
\setlength{\tabcolsep}{3pt}
\renewcommand{\arraystretch}{1.06}
\begin{tabular*}{\linewidth}{@{\extracolsep{\fill}}llrrrrrrr@{}}
\toprule
Record & Representation & $\overline K$ & $\overline\enst$ & $\overline S$ & $\overline F$ & $d_\nabla$ & $\tau_e(8)$ & $\tau_e(24)$ \\
\midrule
FP64 & Supplied & 1.617 & 19.20 & -0.413 & 4.276 & $1.33\!\times\!10^{-7}$ & 0.245 & 0.089 \\
 & SIREN, joint & 1.617 & 19.15 & -0.412 & 4.287 & $2.08\!\times\!10^{-2}$ & 0.245 & 0.092 \\
 & Gaussian, joint & 1.617 & 19.15 & -0.408 & 4.270 & $4.13\!\times\!10^{-2}$ & 0.245 & 0.091 \\
 & SIREN, indep. & 1.616 & 18.72 & -0.387 & 4.213 & $7.03\!\times\!10^{-2}$ & 0.245 & 0.050 \\
 & Gaussian, indep. & 1.616 & 19.00 & -0.393 & 4.221 & $7.69\!\times\!10^{-2}$ & 0.245 & 0.066 \\
\addlinespace[4pt]
FP32 & Supplied & 1.612 & 19.12 & -0.420 & 4.260 & $1.77\!\times\!10^{-7}$ & 0.245 & 0.089 \\
 & SIREN, joint & 1.612 & 19.08 & -0.420 & 4.272 & $2.09\!\times\!10^{-2}$ & 0.245 & 0.091 \\
 & Gaussian, joint & 1.612 & 19.07 & -0.416 & 4.256 & $4.13\!\times\!10^{-2}$ & 0.245 & 0.091 \\
 & SIREN, indep. & 1.611 & 18.65 & -0.395 & 4.196 & $7.07\!\times\!10^{-2}$ & 0.245 & 0.049 \\
 & Gaussian, indep. & 1.611 & 18.92 & -0.401 & 4.206 & $7.70\!\times\!10^{-2}$ & 0.245 & 0.066 \\
\addlinespace[4pt]
FP16 & Supplied & 1.573 & 17.80 & -0.449 & 4.219 & $2.15\!\times\!10^{-3}$ & 0.246 & 0.092 \\
 & SIREN, joint & 1.573 & 17.76 & -0.448 & 4.229 & $2.07\!\times\!10^{-2}$ & 0.246 & 0.094 \\
 & Gaussian, joint & 1.573 & 17.75 & -0.445 & 4.214 & $4.22\!\times\!10^{-2}$ & 0.246 & 0.093 \\
 & SIREN, indep. & 1.572 & 17.33 & -0.424 & 4.168 & $7.21\!\times\!10^{-2}$ & 0.246 & 0.048 \\
 & Gaussian, indep. & 1.572 & 17.61 & -0.430 & 4.164 & $7.90\!\times\!10^{-2}$ & 0.246 & 0.065 \\
\bottomrule
\end{tabular*}
\end{table}

\begin{table}[!ht]
\centering\normalsize
\caption{Absolute diagnostics of supplied records and neural fits: quantization.}
\label{tab:neural-diagnostics-quantization}
\setlength{\tabcolsep}{3pt}
\renewcommand{\arraystretch}{1.06}
\begin{tabular*}{\linewidth}{@{\extracolsep{\fill}}llrrrrrrr@{}}
\toprule
Record & Representation & $\overline K$ & $\overline\enst$ & $\overline S$ & $\overline F$ & $d_\nabla$ & $\tau_e(8)$ & $\tau_e(24)$ \\
\midrule
FP64 & Supplied & 1.617 & 19.20 & -0.413 & 4.276 & $1.33\!\times\!10^{-7}$ & 0.245 & 0.089 \\
 & SIREN, joint & 1.617 & 19.15 & -0.412 & 4.287 & $2.08\!\times\!10^{-2}$ & 0.245 & 0.092 \\
 & Gaussian, joint & 1.617 & 19.15 & -0.408 & 4.270 & $4.13\!\times\!10^{-2}$ & 0.245 & 0.091 \\
 & SIREN, indep. & 1.616 & 18.72 & -0.387 & 4.213 & $7.03\!\times\!10^{-2}$ & 0.245 & 0.050 \\
 & Gaussian, indep. & 1.616 & 19.00 & -0.393 & 4.221 & $7.69\!\times\!10^{-2}$ & 0.245 & 0.066 \\
\addlinespace[4pt]
FP32 & Supplied & 1.617 & 19.20 & -0.413 & 4.276 & $1.34\!\times\!10^{-7}$ & 0.245 & 0.089 \\
 & SIREN, joint & 1.617 & 19.16 & -0.412 & 4.287 & $2.05\!\times\!10^{-2}$ & 0.245 & 0.092 \\
 & Gaussian, joint & 1.617 & 19.15 & -0.408 & 4.270 & $4.12\!\times\!10^{-2}$ & 0.245 & 0.091 \\
 & SIREN, indep. & 1.616 & 18.73 & -0.387 & 4.215 & $7.02\!\times\!10^{-2}$ & 0.245 & 0.050 \\
 & Gaussian, indep. & 1.616 & 19.00 & -0.393 & 4.221 & $7.69\!\times\!10^{-2}$ & 0.245 & 0.066 \\
\addlinespace[4pt]
FP16 & Supplied & 1.617 & 19.20 & -0.413 & 4.276 & $1.01\!\times\!10^{-4}$ & 0.245 & 0.089 \\
 & SIREN, joint & 1.617 & 19.15 & -0.412 & 4.287 & $2.08\!\times\!10^{-2}$ & 0.245 & 0.092 \\
 & Gaussian, joint & 1.617 & 19.15 & -0.408 & 4.270 & $4.12\!\times\!10^{-2}$ & 0.245 & 0.091 \\
 & SIREN, indep. & 1.616 & 18.72 & -0.387 & 4.213 & $7.06\!\times\!10^{-2}$ & 0.245 & 0.049 \\
 & Gaussian, indep. & 1.616 & 19.00 & -0.393 & 4.222 & $7.69\!\times\!10^{-2}$ & 0.245 & 0.066 \\
\addlinespace[4pt]
FP8 & Supplied & 1.615 & 19.19 & -0.410 & 4.261 & $2.44\!\times\!10^{-2}$ & 0.245 & 0.087 \\
 & SIREN, joint & 1.615 & 19.13 & -0.411 & 4.284 & $2.56\!\times\!10^{-2}$ & 0.245 & 0.092 \\
 & Gaussian, joint & 1.615 & 19.13 & -0.408 & 4.266 & $4.36\!\times\!10^{-2}$ & 0.245 & 0.091 \\
 & SIREN, indep. & 1.614 & 18.70 & -0.387 & 4.212 & $7.12\!\times\!10^{-2}$ & 0.245 & 0.050 \\
 & Gaussian, indep. & 1.614 & 18.98 & -0.393 & 4.219 & $7.77\!\times\!10^{-2}$ & 0.245 & 0.066 \\
\addlinespace[4pt]
FP4 & Supplied & 1.609 & 22.82 & -0.108 & 4.002 & $2.59\!\times\!10^{-1}$ & 0.244 & 0.071 \\
 & SIREN, joint & 1.607 & 21.78 & -0.178 & 4.030 & $2.33\!\times\!10^{-1}$ & 0.244 & 0.082 \\
 & Gaussian, joint & 1.607 & 21.79 & -0.158 & 4.099 & $2.30\!\times\!10^{-1}$ & 0.243 & 0.078 \\
 & SIREN, indep. & 1.601 & 19.46 & -0.291 & 3.954 & $1.72\!\times\!10^{-1}$ & 0.244 & 0.032 \\
 & Gaussian, indep. & 1.602 & 19.96 & -0.295 & 4.127 & $1.72\!\times\!10^{-1}$ & 0.244 & 0.050 \\
\bottomrule
\end{tabular*}
\end{table}

\clearpage
\subsection{Comparison and fitting protocol}
\label{app:neural-fitting-protocol}

\paragraph{What is controlled.}
SIRENs share dense transformations across the domain. Gaussian fields assemble
localized, adaptive contributions. These different inductive biases may
preserve different aspects of the same supplied motion.
We examine those differences with the spatial and temporal measures of
Appendix~\ref{app:turbulence-characteristics}.
At each $M$, both families are fitted by velocity MSE to a common stopping
criterion under an approximately matched total parameter budget $B$.
This compares the representations obtained under a specified fitting protocol.
It does not require equal final MSE.

We distribute $B$ as in Appendix~\ref{app:neural-sharing}.
A joint fit uses all parameters in one function $f_\theta(\pos,t)$.
An independent fit divides them nearly equally among the $M$ functions
$f_{\theta_j}(\pos)$.
Both procedures reconstruct the same observed samples.

\paragraph{Coordinates and normalization.}
Each grid index $i$ becomes $2i/M-1$.
Joint models receive the four normalized coordinates in the order $(t,x,y,z)$.
Independent models receive $(x,y,z)$ and use the frame index
to select parameters.
We subtract one scalar mean $\mu$ and divide by one scalar standard deviation
$s_u$ over all velocity components in the movie.
Quantized targets use the corresponding unquantized target's $\mu$ and $s_u$.
The same constants are used for every fitting procedure and network family
on a given target. Exporting the field reverses this transformation.

\paragraph{Loss and sampling.}
Training minimizes~(\ref{eq:app-neural-mse}) in these standardized units.
Every step samples $65{,}536$ space--time grid locations independently and
uniformly, with replacement. All three velocity components contribute equally.
This is one batch for the complete movie, not one batch per independent frame.
For independent fitting, samples are grouped by frame without changing their
weights. No derivative loss, temporal penalty, or equation residual is
included.

\paragraph{Optimization and stopping.}
We use Adam~\citep{kingma2015adam} with $(\beta_1,\beta_2)=(0.9,0.999)$ and $\epsilon=10^{-8}$,
without weight decay. Each cycle contains 20,000 updates with a cosine
learning-rate decay to zero. We retain the fitted parameters, but reset the
Adam state and learning-rate schedule at the start of each cycle.
Table~\ref{tab:neural-learning-rates} gives the initial rates.
The joint filtering runs supplied these settings, which are reused across
targets and fitting procedures without target-specific retuning.

After each cycle, we evaluate MSE over the complete supplied grid.
Let $\mathcal L_c$ be this standardized MSE and $R$ the fixed standardized
velocity range used for that target. We compute
$P_c=10\log_{10}(R^2/\mathcal L_c)$.
Training stops when two consecutive values of $|P_c-P_{c-1}|$, rounded to
four decimals, are below $0.1$ dB.
Otherwise it stops at 30 cycles, or 600,000 updates.
The last cycle supplies the checkpoint.
For independent fits, the criterion uses the complete movie, not a separate
stopping decision for each frame.
Here, \emph{convergence} means this measured plateau, not a demonstrated
global minimum. All runs included here reached the plateau criterion before
the cap.

\begin{table}[!ht]
\centering\normalsize
\caption{Initial learning rates within each restart cycle.}
\label{tab:neural-learning-rates}
\setlength{\tabcolsep}{6pt}
\begin{tabular*}{\linewidth}{@{\extracolsep{\fill}}lrrr@{}}
\toprule
Parameters & $M=32$ & $M=64$ & $M=128$ \\
\midrule
SIREN weights, biases & $6\times10^{-4}$ & $3\times10^{-4}$ & $3\times10^{-4}$ \\
Gaussian centers & $0.024$ & $0.006$ & $0.003$ \\
Gaussian log-diagonal and off-diagonal shape & $0.040$ & $0.010$ & $0.005$ \\
Gaussian velocity coefficients & $0.080$ & $0.020$ & $0.010$ \\
\bottomrule
\end{tabular*}
\end{table}

\newpage
\subsection{Architecture specifications}
\label{app:neural-architecture-details}

\paragraph{SIREN.}
All variants use five hidden sine layers and a linear three-component output.
The first hidden layer applies $\sin(Mh)$ to its affine preactivation $h$.
Later hidden layers apply $\sin(h)$.
Joint networks have constant hidden width $288$, $1156$, or $4630$ for
$M=32,64,128$, respectively.
At $M=32$, independent networks have width 49 for six frames and 50
for the other 26. At $M=64$, 44 frames use width 143 and 20 use width 144.
At $M=128$, all 128 frames use width 408.
Frames receiving either width are assigned by a fixed permutation with seed
$20260912$.

Weights entering the first layer are drawn uniformly from
$[-1/d_{\mathrm{in}},1/d_{\mathrm{in}}]$.
Subsequent layers, including the output, use bounds
$\pm\sqrt{6/d_{\mathrm{in}}}/M$, where $d_{\mathrm{in}}$ is the input width.
Biases start at zero.

\paragraph{Additive Gaussian fields.}
We instantiate the field in Appendix~\ref{app:neural-families} with
$g=9,18,36$ cells per coordinate at $M=32,64,128$.
The normalized cell spacing is $s_g=2/g$.
Joint fitting places three primitives in each cell of a four-dimensional
lattice. Each primitive has four center parameters, ten lower-triangular
shape parameters, and three signed velocity coefficients.
The off-diagonal shape entries allow space--time coupling.
Independent fitting instead uses a separate three-dimensional bank per frame,
with 12 parameters per primitive: three for its center, six for its shape,
and three for its velocity.

A raw center parameter $r$ gives an offset $(s_g/2)\tanh r$ from its cell
center. Positive diagonal shape entries are parameterized by their logarithms.
For minimum-image coordinate displacement $d$, the compact taper is
\[
 w(d)=\left[\max\!\left(1-
       \left(\frac{\|d\|_\infty}{2s_g}\right)^2,0\right)\right]^3.
\]
Contributions are added without normalization or opacity compositing.
Coordinate differences wrap with period two.
This includes time in the joint Gaussian model, an imposed boundary convention
rather than an assertion that the observation window is physically periodic.
SIREN has no corresponding periodic boundary constraint.

Raw centers start uniformly in $[-1,1]$, off-diagonal entries at zero, and
velocity coefficients as Gaussian values with standard deviation $0.01$.
The initial isotropic width is $0.6s_g/3^{1/4}$ for both procedures.
Independent banks distribute primitives as evenly as possible across frames.
Fixed cell assignments are specified by the construction seed $20260912$.

\newpage

\begin{table}[!ht]
\centering\normalsize
\caption{Total learned scalar parameters per movie.}
\label{tab:neural-architectures}
\setlength{\tabcolsep}{6pt}
\renewcommand{\arraystretch}{1.12}
\begin{tabular*}{\linewidth}{@{\extracolsep{\fill}}llrrr@{}}
\toprule
Family & Procedure & $M=32$ & $M=64$ & $M=128$ \\
\midrule
SIREN & Joint & $335{,}235$ & $5{,}359{,}219$ & $85{,}803{,}163$ \\
& Independent & $335{,}254$ & $5{,}358{,}988$ & $85{,}804{,}416$ \\
\addlinespace
Gaussian & Joint & $334{,}611$ & $5{,}353{,}776$ & $85{,}660{,}416$ \\
& Independent & $334{,}608$ & $5{,}353{,}776$ & $85{,}660{,}416$ \\
\bottomrule
\end{tabular*}
\end{table}

All learned parameters use FP32. Within each family, independent counts differ from joint counts by less than
$0.006\%$ because layer widths and primitive numbers are integers.
Across the two families, counts differ by less than $0.2\%$.
The budget increases with $M$, so the filtering comparison changes bandwidth
and capacity together.

\subsection{Evaluation and reproducibility}
\label{app:neural-evaluation}

\paragraph{Common measurements.}
We export each fit at every supplied coordinate and recorded time, restore
physical velocity units, and apply the same spectral derivatives and temporal
estimator as for its target.
All $M$ frames enter the spatial statistics.
Energy spectra and correlation times are displayed through the target cutoff,
while scalar gradient statistics include every mode on the evaluation grid.
We apply no extra filtering or divergence-free projection to the prediction.
This evaluates reconstruction of the supplied movie, not forecasting or
accuracy between recorded coordinates.

\paragraph{Numerical implementation.}
JAX and Optax implement the training loops.
Parameters and training arrays are FP32. Full-grid error sums use binary64
on the host. The stopping evaluator retains the original JAX matrix-product
setting, which can use TF32 on the GPU.
Exported fields instead use the highest FP32 matrix-product precision.
Consequently, exported MSE can differ slightly from the stopping value.
JAX's 64-bit option is enabled for SIREN to retain the original random-sampling
stream, without changing parameter dtype.
Validated batching and compilation changes preserve the model, sampled loss,
and stopping rule. Per-run manifests identify the backend and source hashes.

The main comparisons use initialization index zero for each setting.
SIREN's base random key is $20260713+W$, with joint width $W$ as above.
Gaussian fields use $20260801+M$.
Independent frame initializations use distinct folded keys.

\paragraph{Sensitivity to training randomness.}
The main study uses one training seed per setting, so the observed differences
between joint and independent fits could in principle reflect a favorable
initialization or sequence of minibatches.
We test this possibility on the unquantized $N=1024$ record at $M=64$.
For each network family, we repeat joint and independent fitting with three
seeds. Only the initialization and minibatch sequence change. The data,
architecture, parameter budget, optimizer, and stopping rule remain fixed,
and all fits reach the prescribed plateau.

Table~\ref{tab:neural-seed-sensitivity} shows that seed variation is small
relative to the differences between fitting procedures. For both network
families, every independent fit retains less enstrophy and has shorter
correlation times than every joint fit. The differences in energy spectra and
gradient moments are likewise larger than their variation across seeds.
Thus, on this record, the distinction between joint and independent fitting is
repeatable rather than the result of one favorable training run.

\begin{table}[!ht]
\centering\normalsize
\caption{Physical diagnostics across three training seeds on one unquantized
$N=1024$ record at $M=64$.}
\label{tab:neural-seed-sensitivity}
\setlength{\tabcolsep}{3pt}
\renewcommand{\arraystretch}{1.15}
\begin{tabular*}{\linewidth}{@{\extracolsep{\fill}}lccccc@{}}
\toprule
Fit & Spectrum error (\%) & Enstrophy & $S$ & $F$ & $\langle\tau/\tau_{\rm rec}\rangle_k$ \\
\midrule
Record & 0 & 81.89 & -0.329 & 3.807 & 1 \\
\midrule
SIREN, Joint & $1.70\!\pm\!0.03$ & $64.17\!\pm\!0.22$ & $-0.289\!\pm\!0.002$ & $3.824\!\pm\!0.003$ & $1.116\!\pm\!0.001$ \\
SIREN, Indep. & $3.63\!\pm\!0.01$ & $46.12\!\pm\!0.14$ & $-0.255\!\pm\!0.001$ & $3.725\!\pm\!0.008$ & $0.801\!\pm\!0.001$ \\
Gaussian, Joint & $0.96\!\pm\!0.00$ & $70.75\!\pm\!0.00$ & $-0.364\!\pm\!0.000$ & $4.052\!\pm\!0.001$ & $1.029\!\pm\!0.000$ \\
Gaussian, Indep. & $3.64\!\pm\!0.00$ & $46.03\!\pm\!0.00$ & $-0.383\!\pm\!0.000$ & $4.448\!\pm\!0.004$ & $0.846\!\pm\!0.000$ \\
\bottomrule
\end{tabular*}
\par\smallskip
\begin{minipage}{\linewidth}\normalsize
Entries are means $\pm$ sample standard deviations over three seeds.
The record row gives the values that a faithful fit would reproduce.
Spectrum error is $100\sum_{k=1}^{31}|E_{\rm fit}(k)-E_{\rm rec}(k)|/
\sum_{k=1}^{31}E_{\rm rec}(k)$.
The last column averages the ratio of fitted to recorded correlation times
across $k=3,\ldots,31$, so values above and below one denote correlations that
persist too long and decay too quickly, respectively.
\end{minipage}
\end{table}

\clearpage
\subsection{Spatial accuracy, temporal behavior, and network family}
\label{app:neural-controls}

\paragraph{Comparing fits at matched spatial error.}
On the clean movie, joint fits attain lower spatial error under the main
study's fixed parameter budget. We ask whether their temporal advantage remains when
snapshot accuracy is comparable. On the same unquantized $M=64$ movie used
above, we select joint and independent checkpoints whose full-grid MSEs agree
within $2\%$. 
Across three seeds per family, joint fits have lower error in adjacent-frame
velocity changes in every pair (Table~\ref{tab:matched-trained-error}).
At twice this time lag, independent fits have lower error in every pair.
Thus, similar snapshot accuracy can accompany different accuracy of the
recorded changes, and the preferred procedure depends on the time interval.

\begin{table}[!ht]
\centering\normalsize
\caption{Trained checkpoints at matched spatial error on the unquantized $N=1024$, $M=64$ movie.}
\label{tab:matched-trained-error}
\setlength{\tabcolsep}{4pt}
\begin{tabular*}{\linewidth}{@{\extracolsep{\fill}}llrrrrr@{}}
\toprule
& & \multicolumn{2}{c}{Spatial MSE ($\times10^2$)} & \multicolumn{3}{c}{Change error: Indep./Joint} \\
Family & Seed & Joint & Indep. & $\Delta t$ & $2\Delta t$ & $4\Delta t$ \\
\midrule
SIREN & 0 & 4.544 & 4.463 & 1.358 & 0.939 & 0.933 \\
SIREN & 1 & 4.552 & 4.517 & 1.367 & 0.947 & 0.944 \\
SIREN & 2 & 5.467 & 5.470 & 1.391 & 0.984 & 0.942 \\
Gaussian & 0 & 3.457 & 3.427 & 1.229 & 0.917 & 1.003 \\
Gaussian & 1 & 3.456 & 3.426 & 1.229 & 0.916 & 1.003 \\
Gaussian & 2 & 3.460 & 3.427 & 1.229 & 0.916 & 1.002 \\
\bottomrule
\end{tabular*}
\par\smallskip
\begin{minipage}{\linewidth}\normalsize
Spatial MSE is measured in standardized velocity units over the full grid.
Each pair agrees within 2\%. Change error is the mean squared error in
the velocity difference between two recorded times. Ratios above one favor
joint fitting.
\end{minipage}
\end{table}

\begin{figure}[p]
\centering
\includegraphics[width=\linewidth]{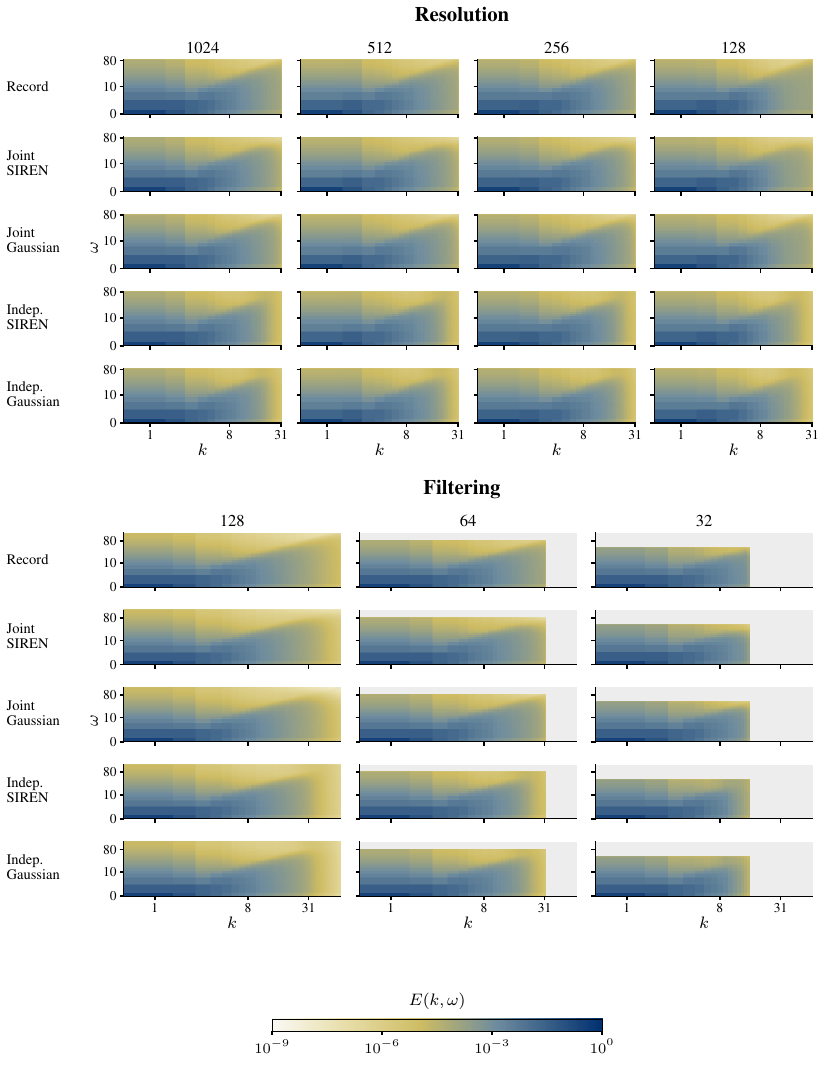}
\caption{\textbf{Joint spectra of supplied records and their neural fits.}
The resolution and filtering comparisons extend
Figure~\ref{fig:joint-versus-independent-neural-representations}. Every column compares a record
with its joint and independent SIREN and Gaussian fits.
Independent fitting broadens small-scale temporal spectra, while joint
fitting tends to narrow them. The maps display absolute energy on a common
color scale, without normalizing each panel. The estimator follows
Figure~\ref{fig:numerical-joint-spectrum}. Gray denotes frequencies outside
the supplied sampling range. The precision and quantization comparisons
continue on the next page.}
\label{fig:neural-joint-spectrum}
\end{figure}

\paragraph{Which spatial scales carry the temporal error?}
A correlation time summarizes how quickly motion at a given spatial scale
changes. The joint spectrum $E(k,\omega)$ shows how its energy is distributed
over temporal frequencies. We compute this spectrum from the recorded
Fourier coefficients over the same observation windows used for the
correlation measurements. Figure~\ref{fig:numerical-joint-spectrum} extends
the numerical comparison, and Figure~\ref{fig:neural-joint-spectrum} compares
each supplied record with its fits. Across the resolution targets,
independent fits spread small-scale energy over higher temporal frequencies,
whereas joint fits tend to concentrate it at lower frequencies.
This pattern remains when we taper the time window to reduce endpoint
leakage and when we use every spatial mode on the clean $M=64$ movie.
The maps locate the changes behind the correlation-time curves:
neural fitting alters both the energy at a spatial scale and how that energy
is distributed in time. The full spectrum complements a single correlation
time, while describing the same second-order temporal information as the
complete correlation function.

\begin{figure}[p]
\centering
\includegraphics[width=\linewidth]{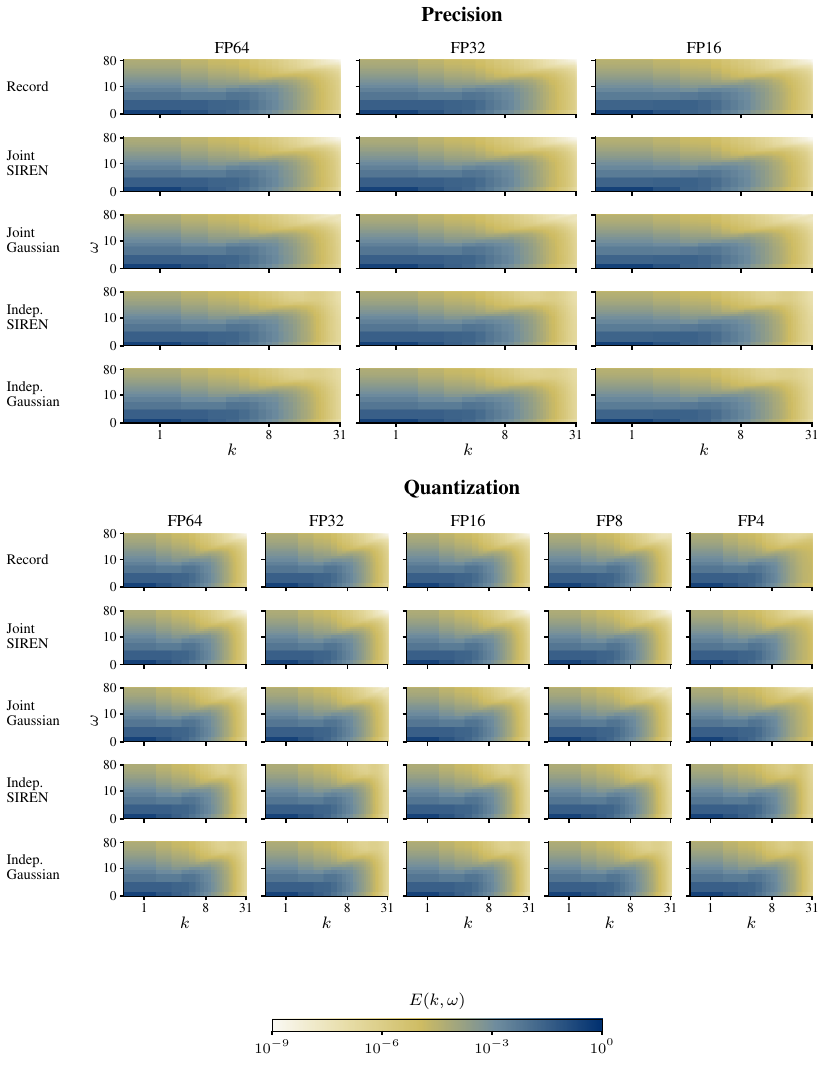}
\par\smallskip
{\normalsize\noindent Figure~\ref{fig:neural-joint-spectrum} (continued).
\textbf{Precision and quantization.} The same estimator, axes, and absolute
color scale compare each record with its fits. Strong quantization and neural
fitting affect both shell energy and its temporal distribution.\par}
\end{figure}

\clearpage
\paragraph{Why do the gradient moments differ between network families?}
We saw that Gaussian fits have more negative gradient skewness and higher flatness than
SIRENs. We ask whether this difference can be explained simply by the spatial
scales retained by each model.

\begin{figure}[!ht]
\centering
\includegraphics[width=\linewidth]{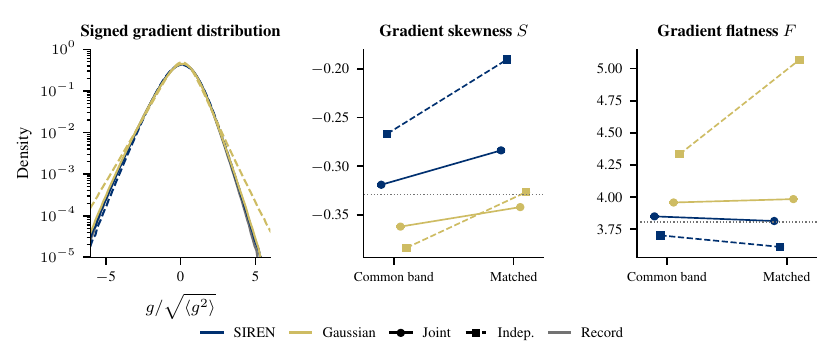}
\caption{\textbf{Matching gradient power does not remove the family difference.}
All fields are restricted to the same spatial band, and their gradient power
is matched to the supplied record shell by shell. Gaussian fits still show
more negative skewness and higher flatness than SIRENs under both fitting
procedures. Points and error bars show means and sample standard deviations
over three training seeds per architecture and fitting procedure.}
\label{fig:gradient-organization}
\end{figure}

To test this, we use the same clean movie with three training seeds for each
architecture and fitting procedure. We restrict all fields to the common shells
$k=1,\ldots,31$ and rescale the longitudinal-gradient power in each shell to
match the record. After this matching, the difference remains:
Gaussian fits still show more negative skewness and higher flatness than
SIRENs under both fitting procedures
(Figure~\ref{fig:gradient-organization}). For independent fits, the matched
flatness is $5.07$ for Gaussians and $3.61$ for SIRENs, compared with $3.81$
in the record.

Thus, the family difference is not simply spectral: even after matching the
gradient power at each spatial scale, Gaussian fits retain more negative
skewness and higher flatness. The architectures therefore differ in how
gradients are organized within those scales, with strong gradients contributing
more prominently in the Gaussian representations.

\end{document}